\documentclass[%
reprint,
superscriptaddress,
prl,
 amsmath,amssymb,
 aps,
]{revtex4-2}

\usepackage[svgnames]{xcolor} 
\usepackage{graphicx,epsfig}
\usepackage{times} 
\usepackage{amssymb,amsmath,multirow,rotate,color}
\usepackage{xcolor}
\usepackage[normalem]{ulem}
\usepackage{comment}

\definecolor{giocolor}{RGB}{0, 150, 100}

\usepackage{float}
\usepackage{lipsum} 
\usepackage{ragged2e}
\usepackage{soul}
\DeclareMathAlphabet\mathbfcal{OMS}{cmsy}{b}{n}

\usepackage{booktabs}

\usepackage[most]{tcolorbox}
\usepackage[svgnames]{xcolor}

\definecolor{Header_color}{HTML}{7500c0} 
\definecolor{Body_color}{HTML}{582F91}

\begin{document}

\title{Dynastic dynamics: Modelling powerful naming choices with stochastic prestige}

\author{Pablo Rosillo-Rodes}
\altaffiliation{These two authors contributed equally. Correspondence: prosillo@ifisc.uib-csic.es, santiagolaot@gmail.com.}
\affiliation{
     Institute for Cross-Disciplinary Physics and Complex Systems IFISC (UIB-CSIC), Campus Universitat de les Illes Balears, E-07122 Palma de Mallorca, Spain
}
\affiliation{
  Computational Story Lab,
  University of Vermont,
  Burlington,
  VT 05405,
  US
}

\affiliation{
Vermont Complex Systems Institute, University of Vermont, USA}

\author{Santiago Lamata-Otín}
\altaffiliation{These two authors contributed equally. Correspondence: prosillo@ifisc.uib-csic.es, santiagolaot@gmail.com.}
\affiliation{Department of Condensed Matter Physics, University of Zaragoza, 50009 Zaragoza, Spain}
\affiliation{GOTHAM lab, Institute of Biocomputation and Physics of Complex Systems (BIFI), University of Zaragoza, 50018 Zaragoza, Spain}
\affiliation{
Vermont Complex Systems Institute, University of Vermont, USA}

\author{David Soriano-Paños}
\affiliation{Departament d’Enginyeria Informàtica i Matemàtiques, Universitat Rovira i Virgili, 43007 Tarragona, Spain}
\affiliation{GOTHAM lab, Institute of Biocomputation and Physics of
Complex Systems (BIFI), University of Zaragoza, 50018 Zaragoza, Spain}

\author{
\firstname{Laurent}
\surname{H\'ebert-Dufresne}
}
\affiliation{
Vermont Complex Systems Institute, University of Vermont, USA}

\affiliation{
   Santa Fe Institute,
   1399 Hyde Park Rd,
   Santa Fe,
   NM 87501,
   US
}

\affiliation{
  Complexity Science Hub,
  Metternichgasse 8,
  1030 Vienna,
  Austria
}

\author{
 \firstname{Peter Sheridan}
 \surname{Dodds}
 }
 \affiliation{
   Computational Story Lab,
   University of Vermont,
   Burlington,
   VT 05405,
   US
 }
\affiliation{
Vermont Complex Systems Institute, University of Vermont, USA}

 \affiliation{
   Santa Fe Institute,
   1399 Hyde Park Rd,
   Santa Fe,
   NM 87501,
   US
 }

\affiliation{
  Complexity Science Hub,
  Metternichgasse 8,
  1030 Vienna,
  Austria
}



\date{\today}


\begin{abstract}
The naming choices of powerful rulers encode millennia of cultural, political, and institutional influences. Consequently, the long-term onomastic dynamics within some of the most powerful dynasties cannot be explained
by simple mechanisms such as frequency-dependent reinforcement or random name reuse. Here, we propose a minimal model that encapsulates these influences in a single highly stochastic variable, prestige, and assume that name selection is driven by the prestige accumulated by previous rulers bearing the same name within each dynasty. Using an extensive dataset spanning ten dynasties, we show that the model reproduces the pronounced inequalities in name frequencies, the long-term persistence of dominant names, and the abrupt rises in popularity observed in historical records. Our results suggest that the naming traditions of powerful rulers preserve institutional memory through reinforcement while remaining sensitive to rare historical events that can reshape the hierarchy of names across generations.
\end{abstract}
\maketitle

\section{Introduction}

How do the powerful choose their names? The naming choices of political and spiritual rulers constitute a paradigmatic case study in cultural evolution~\cite{lieberson2000matter,alford1987naming}. The selection of a regnal or ruling name forms a discrete sequence of institutional choices, shaped by the interplay of power, tradition, and reform~\cite{nau1993,devinne2006,poole1917}. In this sense, the names of the powerful are symbolic assets rather than neutral labels. They encode social memory, institutional continuity, and collective values, transforming each act of naming into a public claim about inheritance, legitimacy, and change~\cite{bourdieu1991language,hahn2003drift,nayef2023,dodds2023}.

Throughout history, rulers have routinely chosen names linked to the land, the divine, and conflict, cultivating associations with strength, perfection, protection, power, and military victory~\cite{nayef2023}. In ancient Egypt, pharaonic names were used to announce royal policies and to imitate the nomenclature of illustrious predecessors~\cite{nayef2023}. In Roman imperial history, names likewise evolved for social and political reasons, often serving to consolidate the civil and military legacy of earlier rulers~\cite{hammond1957,salway1994}. A similar logic appears in the centuries-long sequence of pontificates during which newly elected popes repeatedly adopted names already chosen by their predecessors~\cite{popes_list,nau1993,devinne2006}. Taken together, these examples showcase that the accumulated prestige of a name is a central mechanism behind dynastic naming traditions.

To quantify how past prestige weighs on present choices, we frame ruling names as onomastic time series whose evolution can be studied through the lens of quantitative complex-systems approaches. Such approaches have been used to model the evolution and governance of societies, language, and culture~\cite{kendall1999, turchin2013, roman2017, roman2021, turchin2022, dodds2023,roman2023, piras2024,diazdiaz2024,hebert2025governance,abrams2003modelling,rosillo2023}. 
In particular, quantitative studies of onomastics have examined cultural drift or papal-name prediction, among other phenomena~\cite{bentley2004,newberry2022,baratto2025relationship,antonioni2025complex,lappo2026}. Yet, the long-term dynamics of ruling-name choices remain largely unexplored.

A natural formal language for modeling such repeated choices is provided by urn models, in which options are repeatedly selected from a repertoire that may change over time~\cite{johnson1977urn,pemantle2007survey,mahmoud2008polya}. 
Urn schemes encompass mechanisms ranging from simple random sampling with~\cite{bernoulli1713ars} or without~\cite{pearson1895x,feller1991introduction} replacement to reinforcement processes in which past selections bias future ones~\cite{eggenberger1923statistik,friedman1949simple,blackwell1973ferguson,hill1980strong,muliere2006randomly}, including heterogeneous reinforcement distributions~\cite{fortini2021predictive,sariev2023infinite} and extensions that include innovation through the introduction of new elements in the repertoire~\cite{hoppe1984polya,aldous2006exchangeability,tria2014dynamics}. 
In complex systems, reinforcement is commonly associated with rich-get-richer (RGR) dynamics and the Matthew effect~\cite{merton1968matthew}, providing a minimal mechanism for heavy-tailed distributions and scaling laws across many empirical domains~\cite{yule1925mathematical,zipf1949human,price1976,redner1998popular,gabaix1999zipf,batty2008size,lewens2018}; related principles also appear in cumulative-advantage models~\cite{simon1955class,Zanette2005,rosillo_2026} and growing-network models~\cite{barabasi1999emergence,gomez2004local}. 
However, in regnal naming, reinforcement is not only frequency-based: rare but highly prestigious reigns may abruptly reshape name popularity and disrupt existing name hierarchies.

%

In this work, we introduce a minimal stochastic model formulated within the framework of randomly reinforced urns~\cite{muliere2006randomly} with innovation, in which reinforcement is interpreted as accumulated regnal prestige. Crucially, prestige is modeled through a heterogeneous heavy-tailed distribution unlike most reinforced urn model assuming uniform reinforcements. We validate our model with a broad set of powerful ruling dynasties, including Egyptian pharaohs, Roman emperors, popes, Russian tsars, patriarchs of Constantinople, and monarchs of several regions. Using the calibrated parameters of the model, we show that the role of prestige heterogeneity is more pronounced in non-hereditary dynasties, whereby the establishment of successful lineages is more shaped by the individual performance of the predecessor than by the historical name frequency within the dynasty.

\section{Results}

\subsection{Onomastics of powerful dynasties}

We analyze the name timelines across 10 distinct ruling systems: Egyptian pharaohs~\cite{pharaons_eg_list}, Roman Emperors~\cite{emperors_rom_list}, popes~\cite{popes_list}, Russian tsars~\cite{zars_ru_list}, the patriarchs of Constantinople~\cite{patriarchs_const_list}, Ottoman sultans, Holy Roman emperors, and the monarchs of Spain~\cite{kings_es_list},
Denmark~\cite{kings_dn_list}, and England~\cite{kings_en_list}. Each discrete event in these timelines corresponds to the succession of a new ruler following the death, abdication, or deposition of their predecessor. We show in Fig. \ref{fig:Fig1} the dynastic dynamics for popes (top row), Egyptian pharaohs (middle row), and Roman emperors (bottom row), and we present the remaining datasets in Supplementary Fig. 1 in Supplementary Note 1. 

Fig. \ref{fig:Fig1}a-c displays the temporal dynamics of name abundance, showing that name usage is bursty rather than smooth. Names can remain dormant for long periods before being rapidly reused, generating cascades in their cumulative abundance. 
At the same time, new names continue to enter the repertoire, highlighting the interplay between innovation and reuse.
This balance is summarized by the evolution of the number of distinct names $M(t)$, as we show in the insets of Fig.~\ref{fig:Fig1}d-f. 
At early times, $M(t)$ grows rapidly, indicating a regime dominated by onomastic innovation. 
As the sequences progress, this growth slows down, and the curves bend away from linearity, showing that the introduction of new names becomes less frequent and reuse increasingly dominates the dynamics.
In hereditary monarchies, extended periods of low innovation often reflect dynastic constraints and lineage-based naming traditions (see Supplementary Fig.~1). 
In contrast, the papal sequence reflects the consolidation of an institutional tradition of choosing names already present in the repertoire, leading to a long period of almost uninterrupted reuse until the election of Francis I in 2013.

Finally, in Fig.~\ref{fig:Fig1}g-i we show that bursty reuse produces increasing inequality in name frequencies. The variance of the name-frequency distribution $\mathcal{V}(t)$  grows over time (see top rows), indicating that repeated use progressively concentrates on a small subset of names. 
Consistently, the rank-frequency $\left(r-f\right)$ distributions measured at the end of the simulations display a broad, power-law-like hierarchy, compatible with the data. Here, names are ordered by descending frequency ($r=1$ for the most common), revealing a structure dominated by a few highly frequent names alongside a long tail of rare ones.

\subsection{Modeling rare but prestigious reigns}

\begin{figure*}[t]
  \centering
  \includegraphics[width=1\linewidth]{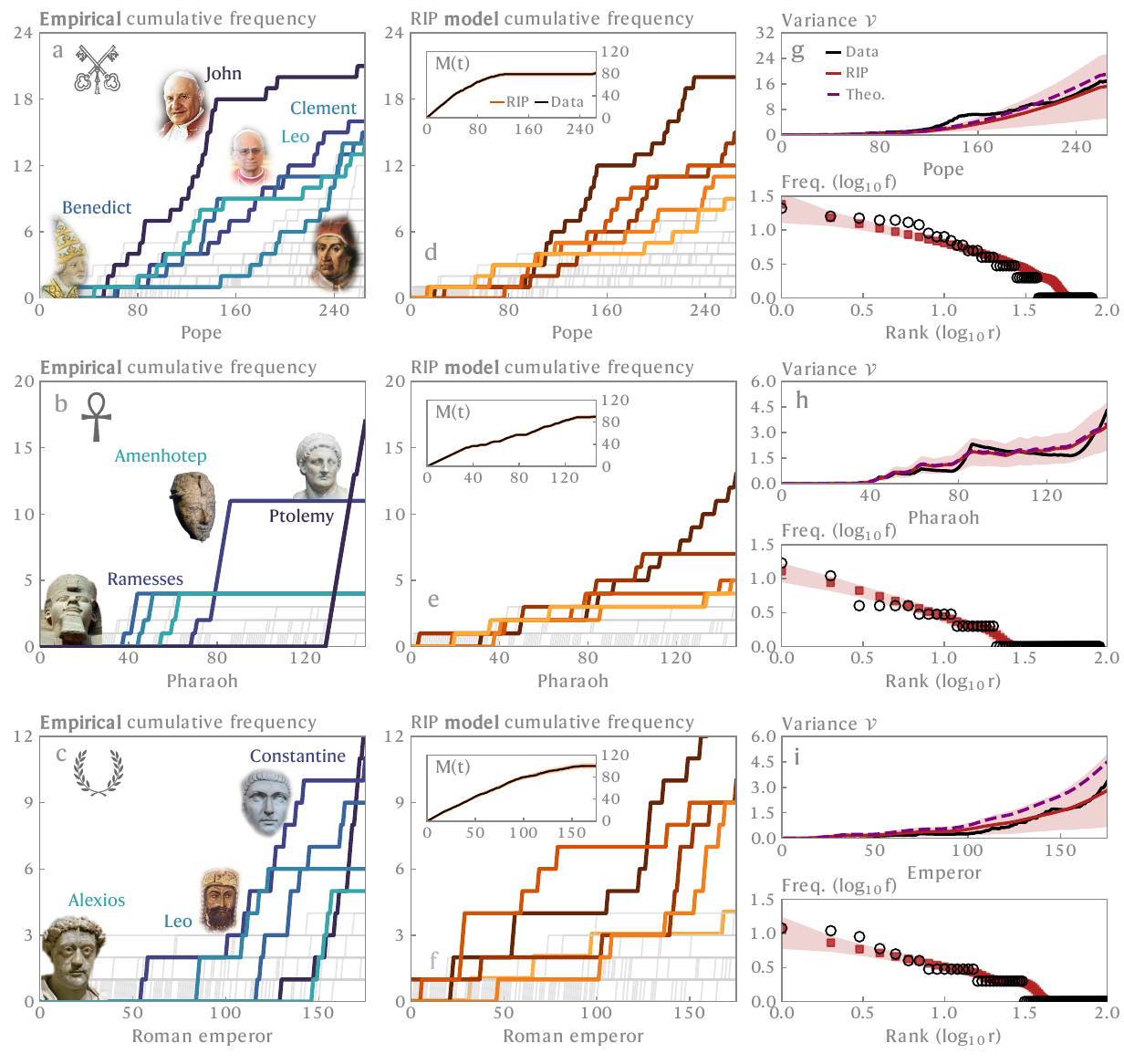}
  \caption{\textbf{Innovation and rare but prestigious reigns explain the onomastics of powerful dynasties}. \textbf{a}-\textbf{c} Number of popes, Egyptian pharaohs, and  Roman emperors respectively, bearing each name across chronologically ordered reigns. \textbf{d}-\textbf{f} Output of a single run of the RIP model.  The insets show the evolution of the number of distinct names $(M(t)$ in bottom rows) over time.
\textbf{g}-\textbf{i} Top: Variance of name frequencies in the data compared with 1000 simulations of the RIP model and with the analytical expression from Eqs.~\eqref{eq:variance_theo}-\eqref{eq:Q_t}. Bottom: rank-frequency distribution of names. The plots show the frequency $f_i$ of each name $i$ as a function of its rank $r_i$, comparing the empirical data with the mean outcome of 1000 simulations of the RIP model $\pm1\sigma$. Note that we standardize the name data by isolating the primary component of authority in each dynastic tradition (see Methods).}
  \label{fig:Fig1}
\end{figure*}

\begin{figure*}[t]
  \centering
  \includegraphics[width=1\linewidth]{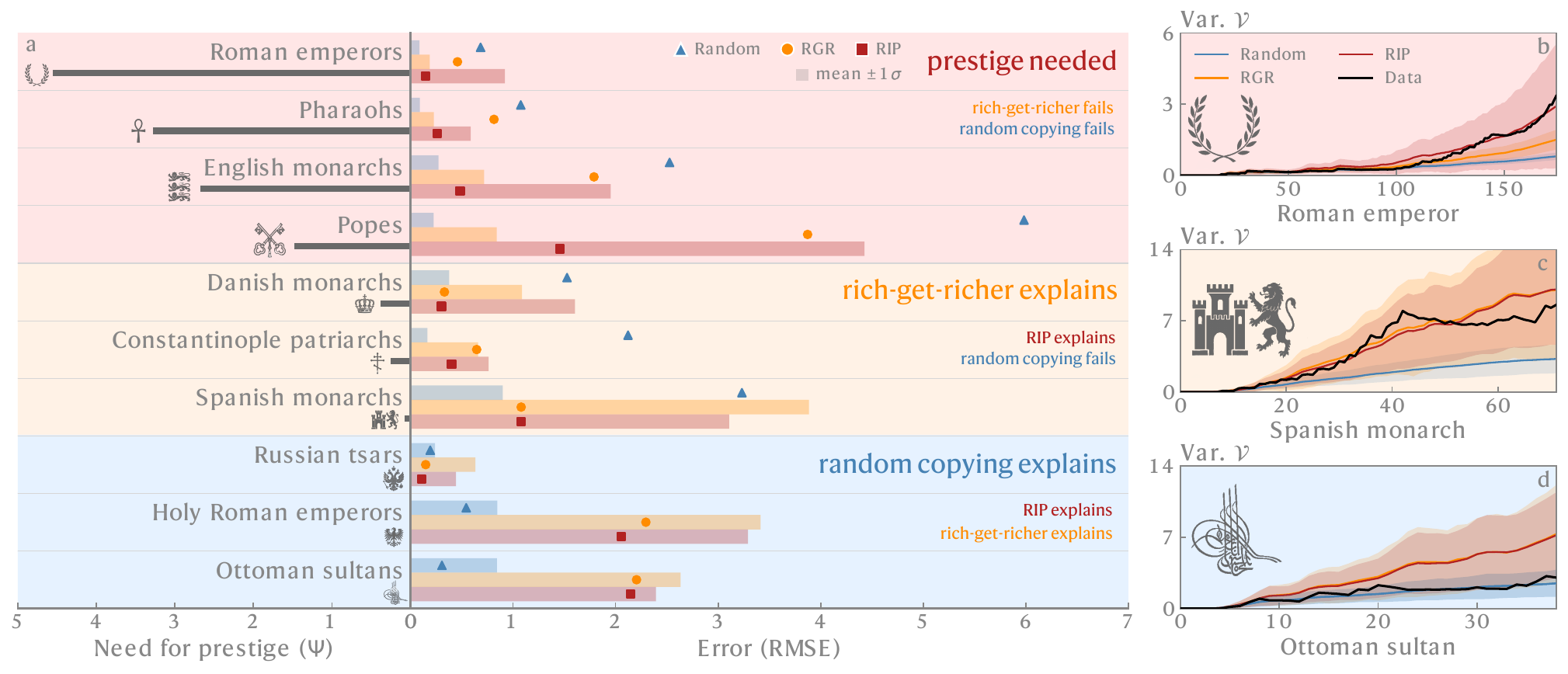}
  \caption{\textbf{Prestige explains what random copying and rich-get-richer dynamics miss about onomastics.} \textbf{a} Root mean square error (RMSE) of the variance of the name distribution for the random, rich-get-richer, and RIP models. Shaded bars represent the RMSE that a trajectory at distance $\pm1\sigma$ of the mean of the trajectory would produce. If a point is inside its corresponding bar, the empirical trajectory is compatible with a random realization of the model at $1\sigma$ confidence. Prestige bars below labels indicate the prestige indicator $\Psi$. \textbf{b}-\textbf{d} Examples of datasets where prestige is or is not required for explaining onomastics. Background colors group the traditions based on the simplest mechanism that accounts for their variance trajectory. Specifically, blue denotes cases explained by the random model (as well as rich-get-richer and RIP), orange refers to dynamics accounted for by rich-get-richer (and RIP), and red highlights instances where prestige (RIP) is explicitly required to explain the trajectory.}
  \label{fig:Fig2}
\end{figure*} 

%


To understand these patterns, we introduce the Reinforcement and Innovation through Prestige  (RIP) model, a minimal stochastic framework assuming that naming choices are determined by the interplay between innovation dynamics and prestige-driven name reuse. Note that, depending on the historical dataset, the meaning of prestige can vary (see Supplementary Note 1). In our model, at each step, with probability $\rho(t)$ a succession event introduces a previously unused name rather than reusing one from the existing repertoire. Otherwise, an already used name is chosen according to the accumulated prestige by its predecessors. After a name is chosen, the prestige of that name is incremented considering a prestige-increment distribution $P\left(K\right)$, with mean given by $\mu=\mathbb{E}[K]$ and variance $\sigma^2=\mathrm{Var}(K)=\mathbb{E}[K^2]-\mu^2$. Note that $\sigma \ll \mu$ ($\sigma \gg \mu$)  produces homogeneous (heterogeneous) individual contributions to the prestige accumulated by each name in the dynasty. 
Note that, when $\sigma^2=0$, prestige becomes proportional to abundance, and this recovers rich-get-richer dynamics. 

In Fig.~\ref{fig:Fig1}d-f we show a realization of the model for each dataset using the innovation function $\rho(t)$ averaged over a centered rolling window of length 3 as input data and the distribution $P\left(K\right)$ fitted as described in Methods. Comparing with the real data in Fig.~\ref{fig:Fig1}a-c, we appreciate that the RIP model captures the abrupt rises in the long-term popularity of individual names. These cascades are then a direct consequence of rare events of prestige gain, since heavy-tailed reinforcement naturally leads to extended periods dominated by a single name.

Furthermore, the top rows in Fig.~\ref{fig:Fig1}g-i show that the RIP model also matches the macroscopic statistics of onomastics. The variance of the simulated name distribution $\overline{\mathcal{V}}(t)$ averaged over 1000 runs of the model aligns closely with the data. Moreover, in Eqs. (\ref{eq:variance_theo})-(\ref{eq:Q_t}) in the Methods section, we have derived an analytical prediction for the time evolution of the variance that fairly reproduces the data. From this analytical treatment we also find that, for constant innovation $\rho(t)=\rho$, the variance is asymptotically bounded only above a critical rate $\rho_c=1/2$. Below this value, $\overline{\mathcal{V}}(t)$ diverges as $t\rightarrow\infty$. Since innovation decays below $\rho_c$ in all datasets (see Supplementary Fig.~3), the continued growth of the variance trajectories in Fig.~\ref{fig:Fig1}g--i is consistent with the model framework.

Finally, the RIP model produces strong inequality in the frequency of names compatible with the one observed in the actual data (see the bottom rows in Fig.~\ref{fig:Fig1}g-i). In the Supplementary Fig. 1 we show the agreement for the remaining datasets.


\subsection{On the necessity of rare but prestigious reigns}

The agreement between the RIP model and the empirical trajectories raises a natural question: is heterogeneous prestige necessary, or can the observed dynamics be explained by simpler reuse mechanisms? To address this, we compare the RIP model with two baselines: a random-choice model, in which reused names are selected uniformly from the existing repertoire, and a rich-get-richer model, in which reused names are selected proportionally to their previous abundance (note that this corresponds to the limit case $\sigma^2=0$ in the RIP model). In Fig. \ref{fig:Fig2}a we plot with a point the root mean square error (RMSE) of the average variance of 1000 simulations of the name distribution with respect to the actual data for the random, rich-get-richer, and RIP models, and with shaded bars we plot the RMSE that a trajectory at distance $\pm1\sigma$ of the average of the simulations would produce. In this sense, if the point is inside the bar, it means that the empirical data is compatible at $1\sigma$ confidence with an arbitrary realization of the model. Supplementary Fig. 2 in Supplementary Note 2 shows an alternative benchmark based on Bayesian information criteria.

In all datasets except Ottoman sultans, Russian tsars, and Holy Roman emperors, the Random model's tendency is not consistent with the variance evolution. In the aforementioned three specific datasets, the dynamics do not require either prestige or a rich-get-richer mechanism to be explained. Indeed, Fig. \ref{fig:Fig2}d shows that the Ottoman sultans trend is better explained by the Random case. Out of the other datasets, the rich-get-richer mechanism with innovation turns out to be as likely as the RIP model to explain Danish monarchs, Constantinople patriarchs, and Spanish monarchs (see Fig. \ref{fig:Fig2}c to see the evolution of the variance for the Spanish monarchs). More importantly, however, for several of the most powerful dynasties in history (Pharaohs, English monarchs, Roman emperors, and popes), prestige is crucially required for explaining dynastic dynamics.  

Notice that in those datasets where the RIP model outperforms the RGR and Random alternatives it is due to the fact that the variance of the distribution of names in the RIP model $\left(\overline{\mathcal V}_{\mathrm{RIP}}\right)$ is substantially larger than what RGR $\left(\overline{\mathcal V}_{\mathrm{RGR}}\right)$ and Random $\left(\overline{\mathcal V}_{\mathrm{R}}\right)$ models can produce (see Fig. \ref{fig:Fig2}b for the Roman emperors case). We can compare the performance of the three models analytically when assuming that the prestige-driven model starts from an initial condition such that the initial prestiges are given by the abundances multiplied by the average of the prestige distribution, $\mu$. As derived in Supplementary Note 3, we obtain that 
\begin{equation}
\overline{\mathcal V}_{\mathrm{RIP}}(t)
\geq
\overline{\mathcal V}_{\mathrm{RGR}}(t)
\geq
\overline{\mathcal V}_{\mathrm{R}}(t).\label{eq:condicion}
\end{equation}
The first inequality follows from the additional heterogeneity induced by random prestige increments, whereas the second follows from the reinforcement of already abundant names in the rich-get-richer mechanism. Random reuse lacks both amplification mechanisms and therefore produces the smallest abundance variance within this comparison.


The condition in Eq. (\ref{eq:condicion}) holds in general. Nonetheless, to make the first inequality explicit, we consider an analytically tractable non-innovation regime, such as the one appearing in papal names for over a millennium (see the inset in Fig. \ref{fig:Fig1}d). In this closed-repertoire limit, the evolution of the variance is fully described by the number of initial name choices $N_0$ and the number of different names $M_0$. Starting from the same initial conditions, we derive in the Supplementary Note 3 that the variance in both the prestige-based and the rich-get-richer models grows quadratically with time (see Eqs. (\ref{eq:v_prestige_noinnovation})-(\ref{eq:v_RGR_noinnovation}) in Methods). From those expressions, the excess heterogeneity generated by the prestige model is given by
\begin{equation}
\overline{\mathcal{V}}_{\mathrm{RIP}}(t) - \overline{\mathcal{V}}_{\mathrm{RGR}}(t) \sim \frac{1}{M_0N_0} \frac{\sigma^2}{\mu^2} (N_0+t)^2. \label{eq:comparison}
\end{equation}
Equation~\eqref{eq:comparison} shows that, although both mechanisms produce quadratic variance growth in the absence of innovation, heterogeneous prestige increments increase the prefactor of this growth. Thus, prestige-driven reinforcement yields a strictly larger variance than rich-get-richer dynamics whenever $\sigma^2>0$.

Based on Eq. (\ref{eq:condicion}), we can define an indicator for the need of prestige to describe the dynamics in terms of the first two moments of the prestige distribution as $\Psi=\sigma^2/\mu^2$. Below each label in  Fig. \ref{fig:Fig2}a, we plot a bar with length $\Psi$. Clearly, those datasets that are better explained by the RIP model are those with a larger value of $\Psi$, meaning that in those cases where the prediction of the RIP model is similar to that of the rich-get-richer or random models are those where the distribution of prestiges that best describes the data does not feature rare events with large prestige.


\section{Discussion}

For a ruler, choosing a name is both a symbolic act and a historical marker. Our analysis indicates that long-term patterns in dynastic names can be described using statistical features found in all kinds of natural systems, being stochastic extreme and rare increments in the accumulated prestige of some lineages necessary to describe many powerful dynasties.



The Reinforcement and Innovation through Prestige  (RIP) model offers a compact quantitative framework to understand the onomastic trends in a wide range of powerful dynasties. With minimal assumptions, it reproduces the marked inequality in name frequencies, the long persistence of preferred names, and the abrupt changes triggered by rare prestige increments. 
Rather than reducing these patterns to standard frequency-based reinforcement, the RIP model defines a prestige-driven reinforcement process in which highly stochastic prestige increments drive naming choices. 

The same formulation also admits a growing network interpretation, where rulers are prestige-endowed nodes, and name lineages are connected components whose growth is driven by the accumulated prestige of their members (see Supplementary Figs.~4 and 5 in Supplementary Note~4). 
In this sense, the model also provides a framework for network growth based on accumulated stochastic fitness, distinct from pure structural rich-get-richer dynamics~\cite{barabasi1999emergence,gomez2004local} or hybrid approaches where fitness acts as a multiplicative factor of connectivity~\cite{bianconi2001bose}.

We analyze several different naming traditions. While both secular and religious naming systems transition from initial linear innovation to periods with zero-innovation plateaus, their underlying mechanisms diverge. Secular monarchies rely on dynastic conservation, creating shorter, intermittent plateaus of name reuse. Conversely, the papacy features a unique, millennial-long institutional plateau of complete stagnation that was only broken by Pope Francis I (see Supplementary Note 1 for more context). These structural variations reveal why some traditions can be modeled by a rich-get-richer framework, whereas other systems like the papacy or Egyptian pharaohs require extreme, event-driven spikes in name prestige to introduce new trends.

Overall, in this work we show how historical sources examined through methods from complex systems can reveal stable structures beneath centuries of political change, cultural evolution and theological shifts. This framework can be applied to other naming traditions, leadership sequences, or institutional lineages, opening a quantitative path for analysing the evolution of social and symbolic conventions. Our results also suggest that long-lived symbolic institutions may evolve through the interplay of steady background dynamics and rare, high-impact events.

Crucially, our work addresses a fundamental open question in cultural evolution: whether random sampling is the primary driver of stochasticity compared to factors like innovation, bursty trends, or decision-making~\cite{lewens2018}. We approach this problem through the lens of onomastics, offering unique empirical insights drawn from the naming data of powerful historical dynasties: random sampling with innovation is not enough to describe the naming dynamics of some of the most powerful dynasties. However, the phenomenon of name prestige, formulated here as stochastic extreme and rare reinforcing events, is not confined to royal naming conventions. Instead, our framework may apply broadly to socio-technical and cultural analyses of innovation, trends, and social norms, aligning with recent investigations into these dynamic processes~\cite{hebert2025self,kolodny2016,centola2018,singhal2020}.

\section{Data availability statement}

The datasets analyzed and codes used during the current study are available in the \texttt{dynastic-dynamics} Github repository, accesible at \href{https://github.com/pablorosillo/dynastic-dynamics/}{https://github.com/pablorosillo/dynastic-dynamics/}.

\section{Acknowledgments} 

We thank Alberto Aparici for valuable insights on the onomastic traditions of the Roman Empire and Teresa Benet-Casaña for guidance on catholic and papal matters. 
We also thank Pedro Jiménez-González, David Sánchez, Natalia Briñas-Pascual, Leah Keating, Federico Malizia, and Sara Oliver-Bonafoux for their comments on a first version of the manuscript.
P.R.-R. and S.L.-O. thank the Vermont Complex Systems Institute for bringing them together in 2025 Fall term, the St. Joseph Oratory in Montreal (CA) for bringing inspiration, and Gema and Izarbe for enjoying the culmination of this manuscript in the \textit{pirineu aragonés}.
P.R.-R. acknowledges support by the Spanish State Research Agency (MICIU/AEI/10.13039/501100011033) and FEDER (UE) under the Mar{\'\i}a de Maeztu project CEX2021-001164-M and
the project COSASTI (PID2024-157493NB-C21 \& PID2024-157493NB-C22). 
S.L.-O. acknowledges financial support from the Departamento de Industria e Innovaci\'on del Gobierno de Arag\'on y Fondo Social Europeo (FENOL group grant E36-23R), from Ministerio de Ciencia e Innovaci\'on (grant PID2020-113582GB-I00), and from Gobierno de Aragón through a doctoral fellowship.
D.S.-P acknowledge Spanish Ministerio de Ciencia e Innovaci\'on (PID2024-158120NB-C21).
We are grateful for
National Science Foundation Award \#2242829
(Science of Online Corpora, Knowledge, and Stories) and award \#2419733,
foundational support from MassMutual, and an anonymous philanthropic gift.

\bibliography{biblio}

\begin{thebibliography}{70}%
\makeatletter
\providecommand \@ifxundefined [1]{%
 \@ifx{#1\undefined}
}%
\providecommand \@ifnum [1]{%
 \ifnum #1\expandafter \@firstoftwo
 \else \expandafter \@secondoftwo
 \fi
}%
\providecommand \@ifx [1]{%
 \ifx #1\expandafter \@firstoftwo
 \else \expandafter \@secondoftwo
 \fi
}%
\providecommand \natexlab [1]{#1}%
\providecommand \enquote  [1]{``#1''}%
\providecommand \bibnamefont  [1]{#1}%
\providecommand \bibfnamefont [1]{#1}%
\providecommand \citenamefont [1]{#1}%
\providecommand \href@noop [0]{\@secondoftwo}%
\providecommand \href [0]{\begingroup \@sanitize@url \@href}%
\providecommand \@href[1]{\@@startlink{#1}\@@href}%
\providecommand \@@href[1]{\endgroup#1\@@endlink}%
\providecommand \@sanitize@url [0]{\catcode `\\12\catcode `\$12\catcode `\&12\catcode `\#12\catcode `\^12\catcode `\_12\catcode `\%12\relax}%
\providecommand \@@startlink[1]{}%
\providecommand \@@endlink[0]{}%
\providecommand \url  [0]{\begingroup\@sanitize@url \@url }%
\providecommand \@url [1]{\endgroup\@href {#1}{\urlprefix }}%
\providecommand \urlprefix  [0]{URL }%
\providecommand \Eprint [0]{\href }%
\providecommand \doibase [0]{https://doi.org/}%
\providecommand \selectlanguage [0]{\@gobble}%
\providecommand \bibinfo  [0]{\@secondoftwo}%
\providecommand \bibfield  [0]{\@secondoftwo}%
\providecommand \translation [1]{[#1]}%
\providecommand \BibitemOpen [0]{}%
\providecommand \bibitemStop [0]{}%
\providecommand \bibitemNoStop [0]{.\EOS\space}%
\providecommand \EOS [0]{\spacefactor3000\relax}%
\providecommand \BibitemShut  [1]{\csname bibitem#1\endcsname}%
\let\auto@bib@innerbib\@empty
\bibitem [{\citenamefont {Lieberson}(2000)}]{lieberson2000matter}%
  \BibitemOpen
  \bibfield  {author} {\bibinfo {author} {\bibfnamefont {S.}~\bibnamefont {Lieberson}},\ }\href@noop {} {\emph {\bibinfo {title} {A matter of taste: How names, fashions, and culture change}}}\ (\bibinfo  {publisher} {Yale University Press},\ \bibinfo {year} {2000})\BibitemShut {NoStop}%
\bibitem [{\citenamefont {Alford}(1987)}]{alford1987naming}%
  \BibitemOpen
  \bibfield  {author} {\bibinfo {author} {\bibfnamefont {R.}~\bibnamefont {Alford}},\ }\bibfield  {title} {\bibinfo {title} {Naming and identity: A cross-cultural study of personal naming practices},\ }\href@noop {} {\bibfield  {journal} {\bibinfo  {journal} {Hraf Press}\ } (\bibinfo {year} {1987})}\BibitemShut {NoStop}%
\bibitem [{\citenamefont {Nau}(1993)}]{nau1993}%
  \BibitemOpen
  \bibfield  {author} {\bibinfo {author} {\bibfnamefont {T.}~\bibnamefont {Nau}},\ }\bibfield  {title} {\bibinfo {title} {The names of the popes},\ }\href@noop {} {\bibfield  {journal} {\bibinfo  {journal} {Onomastica Canadiana}\ }\textbf {\bibinfo {volume} {75}} (\bibinfo {year} {1993})}\BibitemShut {NoStop}%
\bibitem [{\citenamefont {De~Vinne}(2006)}]{devinne2006}%
  \BibitemOpen
  \bibfield  {author} {\bibinfo {author} {\bibfnamefont {C.}~\bibnamefont {De~Vinne}},\ }\bibfield  {title} {\bibinfo {title} {Papal self-naming: Genesis of a tradition},\ }\href@noop {} {\bibfield  {journal} {\bibinfo  {journal} {Onomastica Canadiana}\ }\textbf {\bibinfo {volume} {88}} (\bibinfo {year} {2006})}\BibitemShut {NoStop}%
\bibitem [{\citenamefont {Poole}(1917)}]{poole1917}%
  \BibitemOpen
  \bibfield  {author} {\bibinfo {author} {\bibfnamefont {R.~L.}\ \bibnamefont {Poole}},\ }\bibfield  {title} {\bibinfo {title} {The names and numbers of medieval popes},\ }\href@noop {} {\bibfield  {journal} {\bibinfo  {journal} {The English Historical Review}\ }\textbf {\bibinfo {volume} {32}},\ \bibinfo {pages} {465} (\bibinfo {year} {1917})}\BibitemShut {NoStop}%
\bibitem [{\citenamefont {Bourdieu}(1991)}]{bourdieu1991language}%
  \BibitemOpen
  \bibfield  {author} {\bibinfo {author} {\bibfnamefont {P.}~\bibnamefont {Bourdieu}},\ }\href@noop {} {\emph {\bibinfo {title} {Language and symbolic power}}}\ (\bibinfo  {publisher} {Harvard University Press},\ \bibinfo {year} {1991})\BibitemShut {NoStop}%
\bibitem [{\citenamefont {Hahn}\ and\ \citenamefont {Bentley}(2003)}]{hahn2003drift}%
  \BibitemOpen
  \bibfield  {author} {\bibinfo {author} {\bibfnamefont {M.~W.}\ \bibnamefont {Hahn}}\ and\ \bibinfo {author} {\bibfnamefont {R.~A.}\ \bibnamefont {Bentley}},\ }\bibfield  {title} {\bibinfo {title} {Drift as a mechanism for cultural change: an example from baby names},\ }\href@noop {} {\bibfield  {journal} {\bibinfo  {journal} {Proceedings of the Royal Society of London. Series B: Biological Sciences}\ }\textbf {\bibinfo {volume} {270}},\ \bibinfo {pages} {S120} (\bibinfo {year} {2003})}\BibitemShut {NoStop}%
\bibitem [{\citenamefont {{Heba Nayef and Mohamed El-Nashar}}(2023)}]{nayef2023}%
  \BibitemOpen
  \bibfield  {author} {\bibinfo {author} {\bibnamefont {{Heba Nayef and Mohamed El-Nashar}}},\ }\bibfield  {title} {\bibinfo {title} {{“Towers of Strength and More”: A Thematic Analysis of Royal Titularies in Ancient Egypt}},\ }\bibfield  {journal} {\bibinfo  {journal} {International Journal of Society, Culture \& Language}\ }\href {https://doi.org/10.22034/ijscl.2023.1990883.2952} {10.22034/ijscl.2023.1990883.2952} (\bibinfo {year} {2023})\BibitemShut {NoStop}%
\bibitem [{\citenamefont {{Peter Sheridan Dodds and Joshua R. Minot and Michael V. Arnold and Thayer Alshaabi and Jane Lydia Adams and David Rushing Dewhurst}}(2023)}]{dodds2023}%
  \BibitemOpen
  \bibfield  {author} {\bibinfo {author} {\bibnamefont {{Peter Sheridan Dodds and Joshua R. Minot and Michael V. Arnold and Thayer Alshaabi and Jane Lydia Adams and David Rushing Dewhurst}}},\ }\bibfield  {title} {\bibinfo {title} {{Allotaxonometry and rank-turbulence divergence: a universal instrument for comparing complex systems}},\ }\href {https://doi.org/10.1140/epjds/s13688-023-00400-x} {\bibfield  {journal} {\bibinfo  {journal} {EPJ Data Science}\ }\textbf {\bibinfo {volume} {12}},\ \bibinfo {pages} {37} (\bibinfo {year} {2023})}\BibitemShut {NoStop}%
\bibitem [{\citenamefont {{Mason Hammond}}(1957)}]{hammond1957}%
  \BibitemOpen
  \bibfield  {author} {\bibinfo {author} {\bibnamefont {{Mason Hammond}}},\ }\bibfield  {title} {\bibinfo {title} {{Imperial Elements in the Formula of the Roman Emperors during the First Two and a Half Centuries of the Empire}},\ }\href {https://doi.org/10.2307/4238646} {\bibfield  {journal} {\bibinfo  {journal} {Memoirs of the American Academy in Rome}\ }\textbf {\bibinfo {volume} {25}},\ \bibinfo {pages} {17} (\bibinfo {year} {1957})}\BibitemShut {NoStop}%
\bibitem [{\citenamefont {{Benet Salway}}(1994)}]{salway1994}%
  \BibitemOpen
  \bibfield  {author} {\bibinfo {author} {\bibnamefont {{Benet Salway}}},\ }\bibfield  {title} {\bibinfo {title} {{What's in a Name? A Survey of Roman Onomastic Practice from c. 700 B.C. to A.D. 700}},\ }\href {https://doi.org/10.2307/300873} {\bibfield  {journal} {\bibinfo  {journal} {The Journal of Roman Studies}\ }\textbf {\bibinfo {volume} {84}},\ \bibinfo {pages} {124} (\bibinfo {year} {1994})}\BibitemShut {NoStop}%
\bibitem [{pop(1911)}]{popes_list}%
  \BibitemOpen
  \href@noop {} {\bibinfo {title} {{The List of Popes}}},\ \bibinfo {howpublished} {\url{http://www.newadvent.org/cathen/12272b.htm}} (\bibinfo {year} {1911}),\ \bibinfo {note} {in \textit{The Catholic Encyclopedia}. New York: Robert Appleton Company. Retrieved online October 31, 2025.}\BibitemShut {Stop}%
\bibitem [{\citenamefont {Kendall}\ \emph {et~al.}(1999)\citenamefont {Kendall}, \citenamefont {Briggs}, \citenamefont {Murdoch}, \citenamefont {Turchin}, \citenamefont {Ellner}, \citenamefont {McCauley}, \citenamefont {Nisbet},\ and\ \citenamefont {Wood}}]{kendall1999}%
  \BibitemOpen
  \bibfield  {author} {\bibinfo {author} {\bibfnamefont {B.~E.}\ \bibnamefont {Kendall}}, \bibinfo {author} {\bibfnamefont {C.~J.}\ \bibnamefont {Briggs}}, \bibinfo {author} {\bibfnamefont {W.~W.}\ \bibnamefont {Murdoch}}, \bibinfo {author} {\bibfnamefont {P.}~\bibnamefont {Turchin}}, \bibinfo {author} {\bibfnamefont {S.~P.}\ \bibnamefont {Ellner}}, \bibinfo {author} {\bibfnamefont {E.}~\bibnamefont {McCauley}}, \bibinfo {author} {\bibfnamefont {R.~M.}\ \bibnamefont {Nisbet}},\ and\ \bibinfo {author} {\bibfnamefont {S.~N.}\ \bibnamefont {Wood}},\ }\bibfield  {title} {\bibinfo {title} {Why do populations cycle? a synthesis of statistical and mechanistic modeling approaches},\ }\href@noop {} {\bibfield  {journal} {\bibinfo  {journal} {Ecology}\ }\textbf {\bibinfo {volume} {80}},\ \bibinfo {pages} {1789} (\bibinfo {year} {1999})}\BibitemShut {NoStop}%
\bibitem [{\citenamefont {Turchin}\ \emph {et~al.}(2013)\citenamefont {Turchin}, \citenamefont {Currie}, \citenamefont {Turner},\ and\ \citenamefont {Gavrilets}}]{turchin2013}%
  \BibitemOpen
  \bibfield  {author} {\bibinfo {author} {\bibfnamefont {P.}~\bibnamefont {Turchin}}, \bibinfo {author} {\bibfnamefont {T.~E.}\ \bibnamefont {Currie}}, \bibinfo {author} {\bibfnamefont {E.~A.~L.}\ \bibnamefont {Turner}},\ and\ \bibinfo {author} {\bibfnamefont {S.}~\bibnamefont {Gavrilets}},\ }\bibfield  {title} {\bibinfo {title} {War, space, and the evolution of old world complex societies},\ }\href {https://doi.org/10.1073/pnas.1308825110} {\bibfield  {journal} {\bibinfo  {journal} {Proceedings of the National Academy of Sciences}\ }\textbf {\bibinfo {volume} {110}},\ \bibinfo {pages} {16384} (\bibinfo {year} {2013})}\BibitemShut {NoStop}%
\bibitem [{\citenamefont {Roman}\ \emph {et~al.}(2017)\citenamefont {Roman}, \citenamefont {Bullock},\ and\ \citenamefont {Brede}}]{roman2017}%
  \BibitemOpen
  \bibfield  {author} {\bibinfo {author} {\bibfnamefont {S.}~\bibnamefont {Roman}}, \bibinfo {author} {\bibfnamefont {S.}~\bibnamefont {Bullock}},\ and\ \bibinfo {author} {\bibfnamefont {M.}~\bibnamefont {Brede}},\ }\bibfield  {title} {\bibinfo {title} {{Coupled Societies are More Robust Against Collapse: A Hypothetical Look at Easter Island}},\ }\href@noop {} {\bibfield  {journal} {\bibinfo  {journal} {Ecological Economics}\ }\textbf {\bibinfo {volume} {132}},\ \bibinfo {pages} {264} (\bibinfo {year} {2017})}\BibitemShut {NoStop}%
\bibitem [{\citenamefont {Roman}(2021)}]{roman2021}%
  \BibitemOpen
  \bibfield  {author} {\bibinfo {author} {\bibfnamefont {S.}~\bibnamefont {Roman}},\ }\bibfield  {title} {\bibinfo {title} {{Historical dynamics of the Chinese dynasties}},\ }\href {https://doi.org/10.1016/j.heliyon.2021.e07293} {\bibfield  {journal} {\bibinfo  {journal} {Heliyon}\ }\textbf {\bibinfo {volume} {7}},\ \bibinfo {pages} {e07293} (\bibinfo {year} {2021})}\BibitemShut {NoStop}%
\bibitem [{\citenamefont {Turchin}\ \emph {et~al.}(2022)\citenamefont {Turchin}, \citenamefont {Whitehouse}, \citenamefont {Gavrilets}, \citenamefont {Hoyer}, \citenamefont {François}, \citenamefont {Bennett}, \citenamefont {Feeney}, \citenamefont {Peregrine}, \citenamefont {Feinman}, \citenamefont {Korotayev}, \citenamefont {Kradin}, \citenamefont {Levine}, \citenamefont {Reddish}, \citenamefont {Cioni}, \citenamefont {Wacziarg}, \citenamefont {Mendel-Gleason},\ and\ \citenamefont {Benam}}]{turchin2022}%
  \BibitemOpen
  \bibfield  {author} {\bibinfo {author} {\bibfnamefont {P.}~\bibnamefont {Turchin}}, \bibinfo {author} {\bibfnamefont {H.}~\bibnamefont {Whitehouse}}, \bibinfo {author} {\bibfnamefont {S.}~\bibnamefont {Gavrilets}}, \bibinfo {author} {\bibfnamefont {D.}~\bibnamefont {Hoyer}}, \bibinfo {author} {\bibfnamefont {P.}~\bibnamefont {François}}, \bibinfo {author} {\bibfnamefont {J.~S.}\ \bibnamefont {Bennett}}, \bibinfo {author} {\bibfnamefont {K.~C.}\ \bibnamefont {Feeney}}, \bibinfo {author} {\bibfnamefont {P.}~\bibnamefont {Peregrine}}, \bibinfo {author} {\bibfnamefont {G.}~\bibnamefont {Feinman}}, \bibinfo {author} {\bibfnamefont {A.}~\bibnamefont {Korotayev}}, \bibinfo {author} {\bibfnamefont {N.}~\bibnamefont {Kradin}}, \bibinfo {author} {\bibfnamefont {J.}~\bibnamefont {Levine}}, \bibinfo {author} {\bibfnamefont {J.}~\bibnamefont {Reddish}}, \bibinfo {author} {\bibfnamefont {E.}~\bibnamefont {Cioni}}, \bibinfo {author} {\bibfnamefont {R.}~\bibnamefont {Wacziarg}}, \bibinfo {author} {\bibfnamefont
  {G.}~\bibnamefont {Mendel-Gleason}},\ and\ \bibinfo {author} {\bibfnamefont {M.}~\bibnamefont {Benam}},\ }\bibfield  {title} {\bibinfo {title} {Disentangling the evolutionary drivers of social complexity: A comprehensive test of hypotheses},\ }\href {https://doi.org/10.1126/sciadv.abn3517} {\bibfield  {journal} {\bibinfo  {journal} {Science Advances}\ }\textbf {\bibinfo {volume} {8}},\ \bibinfo {pages} {eabn3517} (\bibinfo {year} {2022})}\BibitemShut {NoStop}%
\bibitem [{\citenamefont {{Roman, S. and Bertolotti, F.}}(2023)}]{roman2023}%
  \BibitemOpen
  \bibfield  {author} {\bibinfo {author} {\bibnamefont {{Roman, S. and Bertolotti, F.}}},\ }\bibfield  {title} {\bibinfo {title} {{Global history, the emergence of chaos and inducing sustainability in networks of socio-ecological systems}},\ }\href {https://doi.org/10.1371/journal.pone.0293391} {\bibfield  {journal} {\bibinfo  {journal} {PLOS ONE}\ }\textbf {\bibinfo {volume} {18}},\ \bibinfo {pages} {e0293391} (\bibinfo {year} {2023})}\BibitemShut {NoStop}%
\bibitem [{\citenamefont {Piras}\ and\ \citenamefont {Bertolotti}(2024)}]{piras2024}%
  \BibitemOpen
  \bibfield  {author} {\bibinfo {author} {\bibfnamefont {A.}~\bibnamefont {Piras}}\ and\ \bibinfo {author} {\bibfnamefont {F.}~\bibnamefont {Bertolotti}},\ }\bibfield  {title} {\bibinfo {title} {How risk preferences shape city-state success: An agent-based model of resource management},\ }in\ \href@noop {} {\emph {\bibinfo {booktitle} {WOA 2024: 25th Workshop "From Objects to Agents"}}}\ (\bibinfo {address} {Forte di Bard, Aosta, Italy},\ \bibinfo {year} {2024})\BibitemShut {NoStop}%
\bibitem [{\citenamefont {Diaz-Diaz}\ \emph {et~al.}(2024)\citenamefont {Diaz-Diaz}, \citenamefont {Bartesaghi},\ and\ \citenamefont {Estrada}}]{diazdiaz2024}%
  \BibitemOpen
  \bibfield  {author} {\bibinfo {author} {\bibfnamefont {F.}~\bibnamefont {Diaz-Diaz}}, \bibinfo {author} {\bibfnamefont {P.}~\bibnamefont {Bartesaghi}},\ and\ \bibinfo {author} {\bibfnamefont {E.}~\bibnamefont {Estrada}},\ }\bibfield  {title} {\bibinfo {title} {Mathematical modeling of local balance in signed networks and its applications to global international analysis},\ }\href@noop {} {\bibfield  {journal} {\bibinfo  {journal} {Journal of Applied Mathematics and Computing}\ }\textbf {\bibinfo {volume} {70}},\ \bibinfo {pages} {6195} (\bibinfo {year} {2024})}\BibitemShut {NoStop}%
\bibitem [{\citenamefont {H{\'e}bert-Dufresne}\ \emph {et~al.}(2025{\natexlab{a}})\citenamefont {H{\'e}bert-Dufresne}, \citenamefont {Landry}, \citenamefont {Lovato}, \citenamefont {St-Onge}, \citenamefont {Young}, \citenamefont {Couture-M{\'e}nard}, \citenamefont {Bernatchez}, \citenamefont {Choquette},\ and\ \citenamefont {Cohen}}]{hebert2025governance}%
  \BibitemOpen
  \bibfield  {author} {\bibinfo {author} {\bibfnamefont {L.}~\bibnamefont {H{\'e}bert-Dufresne}}, \bibinfo {author} {\bibfnamefont {N.~W.}\ \bibnamefont {Landry}}, \bibinfo {author} {\bibfnamefont {J.}~\bibnamefont {Lovato}}, \bibinfo {author} {\bibfnamefont {J.}~\bibnamefont {St-Onge}}, \bibinfo {author} {\bibfnamefont {J.-G.}\ \bibnamefont {Young}}, \bibinfo {author} {\bibfnamefont {M.-{\`E}.}\ \bibnamefont {Couture-M{\'e}nard}}, \bibinfo {author} {\bibfnamefont {S.}~\bibnamefont {Bernatchez}}, \bibinfo {author} {\bibfnamefont {C.}~\bibnamefont {Choquette}},\ and\ \bibinfo {author} {\bibfnamefont {A.~A.}\ \bibnamefont {Cohen}},\ }\bibfield  {title} {\bibinfo {title} {Governance as a complex, networked, democratic, satisfiability problem},\ }\href@noop {} {\bibfield  {journal} {\bibinfo  {journal} {npj Complexity}\ }\textbf {\bibinfo {volume} {2}},\ \bibinfo {pages} {14} (\bibinfo {year} {2025}{\natexlab{a}})}\BibitemShut {NoStop}%
\bibitem [{\citenamefont {Abrams}\ and\ \citenamefont {Strogatz}(2003)}]{abrams2003modelling}%
  \BibitemOpen
  \bibfield  {author} {\bibinfo {author} {\bibfnamefont {D.~M.}\ \bibnamefont {Abrams}}\ and\ \bibinfo {author} {\bibfnamefont {S.~H.}\ \bibnamefont {Strogatz}},\ }\bibfield  {title} {\bibinfo {title} {Modelling the dynamics of language death},\ }\href@noop {} {\bibfield  {journal} {\bibinfo  {journal} {Nature}\ }\textbf {\bibinfo {volume} {424}},\ \bibinfo {pages} {900} (\bibinfo {year} {2003})}\BibitemShut {NoStop}%
\bibitem [{\citenamefont {Rosillo-Rodes}\ \emph {et~al.}(2023)\citenamefont {Rosillo-Rodes}, \citenamefont {San~Miguel},\ and\ \citenamefont {Sánchez}}]{rosillo2023}%
  \BibitemOpen
  \bibfield  {author} {\bibinfo {author} {\bibfnamefont {P.}~\bibnamefont {Rosillo-Rodes}}, \bibinfo {author} {\bibfnamefont {M.}~\bibnamefont {San~Miguel}},\ and\ \bibinfo {author} {\bibfnamefont {D.}~\bibnamefont {Sánchez}},\ }\bibfield  {title} {\bibinfo {title} {Modeling language ideologies for the dynamics of languages in contact},\ }\href {https://doi.org/10.1063/5.0166636} {\bibfield  {journal} {\bibinfo  {journal} {Chaos}\ }\textbf {\bibinfo {volume} {33}},\ \bibinfo {pages} {113117} (\bibinfo {year} {2023})}\BibitemShut {NoStop}%
\bibitem [{\citenamefont {{R. Alexander Bentley and Matthew W. Hahn and Stephen J. Shennan}}(2004)}]{bentley2004}%
  \BibitemOpen
  \bibfield  {author} {\bibinfo {author} {\bibnamefont {{R. Alexander Bentley and Matthew W. Hahn and Stephen J. Shennan}}},\ }\bibfield  {title} {\bibinfo {title} {{Random drift and culture change}},\ }\href {https://doi.org/10.1098/rspb.2004.2746} {\bibfield  {journal} {\bibinfo  {journal} {Proceedings of the Royal Society of London. Series B: Biological Sciences}\ }\textbf {\bibinfo {volume} {271}},\ \bibinfo {pages} {1443} (\bibinfo {year} {2004})}\BibitemShut {NoStop}%
\bibitem [{\citenamefont {Newberry}\ and\ \citenamefont {Plotkin}(2022)}]{newberry2022}%
  \BibitemOpen
  \bibfield  {author} {\bibinfo {author} {\bibfnamefont {M.~G.}\ \bibnamefont {Newberry}}\ and\ \bibinfo {author} {\bibfnamefont {J.~B.}\ \bibnamefont {Plotkin}},\ }\bibfield  {title} {\bibinfo {title} {Measuring frequency-dependent selection in culture},\ }\href@noop {} {\bibfield  {journal} {\bibinfo  {journal} {Nature Human Behaviour}\ }\textbf {\bibinfo {volume} {6}},\ \bibinfo {pages} {1048} (\bibinfo {year} {2022})}\BibitemShut {NoStop}%
\bibitem [{\citenamefont {Baratto}\ \emph {et~al.}(2025)\citenamefont {Baratto}, \citenamefont {Casanovas}, \citenamefont {Decostanzi}, \citenamefont {Borges}, \citenamefont {Alcal{\'a}}, \citenamefont {Stanzani}, \citenamefont {Antonioni}, \citenamefont {Iacopini}, \citenamefont {Fiorentin},\ and\ \citenamefont {Valdano}}]{baratto2025relationship}%
  \BibitemOpen
  \bibfield  {author} {\bibinfo {author} {\bibfnamefont {M.}~\bibnamefont {Baratto}}, \bibinfo {author} {\bibfnamefont {I.}~\bibnamefont {Casanovas}}, \bibinfo {author} {\bibfnamefont {I.}~\bibnamefont {Decostanzi}}, \bibinfo {author} {\bibfnamefont {H.~M.}\ \bibnamefont {Borges}}, \bibinfo {author} {\bibfnamefont {S.~M.}\ \bibnamefont {Alcal{\'a}}}, \bibinfo {author} {\bibfnamefont {I.}~\bibnamefont {Stanzani}}, \bibinfo {author} {\bibfnamefont {A.}~\bibnamefont {Antonioni}}, \bibinfo {author} {\bibfnamefont {I.}~\bibnamefont {Iacopini}}, \bibinfo {author} {\bibfnamefont {M.~R.}\ \bibnamefont {Fiorentin}},\ and\ \bibinfo {author} {\bibfnamefont {E.}~\bibnamefont {Valdano}},\ }\bibfield  {title} {\bibinfo {title} {The relationship between episcopal genealogy and ideology in the roman catholic church},\ }\href@noop {} {\bibfield  {journal} {\bibinfo  {journal} {arXiv preprint arXiv:2506.22108}\ } (\bibinfo {year} {2025})}\BibitemShut {NoStop}%
\bibitem [{\citenamefont {Antonioni}\ \emph {et~al.}(2025)\citenamefont {Antonioni}, \citenamefont {Fiorentin},\ and\ \citenamefont {Valdano}}]{antonioni2025complex}%
  \BibitemOpen
  \bibfield  {author} {\bibinfo {author} {\bibfnamefont {A.}~\bibnamefont {Antonioni}}, \bibinfo {author} {\bibfnamefont {M.~R.}\ \bibnamefont {Fiorentin}},\ and\ \bibinfo {author} {\bibfnamefont {E.}~\bibnamefont {Valdano}},\ }\bibfield  {title} {\bibinfo {title} {{Complex totopapa: predicting the successor to pope Francis}},\ }\href@noop {} {\bibfield  {journal} {\bibinfo  {journal} {arXiv preprint arXiv:2505.01553}\ } (\bibinfo {year} {2025})}\BibitemShut {NoStop}%
\bibitem [{\citenamefont {{Egor Lappo and Noah A. Rosenberg}}(2026)}]{lappo2026}%
  \BibitemOpen
  \bibfield  {author} {\bibinfo {author} {\bibnamefont {{Egor Lappo and Noah A. Rosenberg}}},\ }\href {https://doi.org/10.48550/arXiv.2605.07028} {\bibinfo {title} {{Quo nomine vis vocari? A random-copying model explains the temporal sequence of papal names}}} (\bibinfo {year} {2026}),\ \bibinfo {note} {arXiv preprint arXiv:2605.07028},\ \Eprint {https://arxiv.org/abs/2605.07028} {arXiv:2605.07028 [q-bio.PE]} \BibitemShut {NoStop}%
\bibitem [{\citenamefont {{N. L. Johnson and S. Kotz}}(1977)}]{johnson1977urn}%
  \BibitemOpen
  \bibfield  {author} {\bibinfo {author} {\bibnamefont {{N. L. Johnson and S. Kotz}}},\ }\href@noop {} {\emph {\bibinfo {title} {{Urn Models and Their Application: An Approach to Modern Discrete Probability Theory}}}}\ (\bibinfo  {publisher} {Wiley},\ \bibinfo {year} {1977})\ p.\ \bibinfo {pages} {402}\BibitemShut {NoStop}%
\bibitem [{\citenamefont {{R. Pemantle}}(2007)}]{pemantle2007survey}%
  \BibitemOpen
  \bibfield  {author} {\bibinfo {author} {\bibnamefont {{R. Pemantle}}},\ }\bibfield  {title} {\bibinfo {title} {{A survey of random processes with reinforcement}},\ }\href {https://doi.org/10.1214/07-PS094} {\bibfield  {journal} {\bibinfo  {journal} {Probability Surveys}\ }\textbf {\bibinfo {volume} {4}},\ \bibinfo {pages} {1} (\bibinfo {year} {2007})}\BibitemShut {NoStop}%
\bibitem [{\citenamefont {Mahmoud}(2008)}]{mahmoud2008polya}%
  \BibitemOpen
  \bibfield  {author} {\bibinfo {author} {\bibfnamefont {H.}~\bibnamefont {Mahmoud}},\ }\href@noop {} {\emph {\bibinfo {title} {P{\'o}lya urn models}}}\ (\bibinfo  {publisher} {Chapman and Hall/CRC},\ \bibinfo {year} {2008})\BibitemShut {NoStop}%
\bibitem [{\citenamefont {Bernoulli}(1713)}]{bernoulli1713ars}%
  \BibitemOpen
  \bibfield  {author} {\bibinfo {author} {\bibfnamefont {J.}~\bibnamefont {Bernoulli}},\ }\href@noop {} {\emph {\bibinfo {title} {Ars conjectandi, opus posthumum: accedit tractatus de seriebus infinitis, et epistola Gallice scripta de ludo pil{\ae} reticularis}}}\ (\bibinfo  {publisher} {Impensis Thurnisiorum Fratrum},\ \bibinfo {year} {1713})\BibitemShut {NoStop}%
\bibitem [{\citenamefont {Pearson}(1895)}]{pearson1895x}%
  \BibitemOpen
  \bibfield  {author} {\bibinfo {author} {\bibfnamefont {K.}~\bibnamefont {Pearson}},\ }\bibfield  {title} {\bibinfo {title} {X. contributions to the mathematical theory of evolution.—ii. skew variation in homogeneous material},\ }\href@noop {} {\bibfield  {journal} {\bibinfo  {journal} {Philosophical Transactions of the Royal Society of London.(A.)}\ ,\ \bibinfo {pages} {343}} (\bibinfo {year} {1895})}\BibitemShut {NoStop}%
\bibitem [{\citenamefont {Feller}(1991)}]{feller1991introduction}%
  \BibitemOpen
  \bibfield  {author} {\bibinfo {author} {\bibfnamefont {W.}~\bibnamefont {Feller}},\ }\href@noop {} {\emph {\bibinfo {title} {An introduction to probability theory and its applications}}},\ Vol.~\bibinfo {volume} {2}\ (\bibinfo  {publisher} {John Wiley \& Sons},\ \bibinfo {year} {1991})\BibitemShut {NoStop}%
\bibitem [{\citenamefont {Eggenberger}\ and\ \citenamefont {P{\'o}lya}(1923)}]{eggenberger1923statistik}%
  \BibitemOpen
  \bibfield  {author} {\bibinfo {author} {\bibfnamefont {F.}~\bibnamefont {Eggenberger}}\ and\ \bibinfo {author} {\bibfnamefont {G.}~\bibnamefont {P{\'o}lya}},\ }\bibfield  {title} {\bibinfo {title} {{\"U}ber die statistik verketteter vorg{\"a}nge},\ }\href@noop {} {\bibfield  {journal} {\bibinfo  {journal} {ZAMM-Journal of Applied Mathematics and Mechanics/Zeitschrift f{\"u}r Angewandte Mathematik und Mechanik}\ }\textbf {\bibinfo {volume} {3}},\ \bibinfo {pages} {279} (\bibinfo {year} {1923})}\BibitemShut {NoStop}%
\bibitem [{\citenamefont {Friedman}(1949)}]{friedman1949simple}%
  \BibitemOpen
  \bibfield  {author} {\bibinfo {author} {\bibfnamefont {B.}~\bibnamefont {Friedman}},\ }\bibfield  {title} {\bibinfo {title} {A simple urn model},\ }\href@noop {} {\bibfield  {journal} {\bibinfo  {journal} {Communications on Pure and Applied Mathematics}\ }\textbf {\bibinfo {volume} {2}},\ \bibinfo {pages} {59} (\bibinfo {year} {1949})}\BibitemShut {NoStop}%
\bibitem [{\citenamefont {Blackwell}\ and\ \citenamefont {MacQueen}(1973)}]{blackwell1973ferguson}%
  \BibitemOpen
  \bibfield  {author} {\bibinfo {author} {\bibfnamefont {D.}~\bibnamefont {Blackwell}}\ and\ \bibinfo {author} {\bibfnamefont {J.~B.}\ \bibnamefont {MacQueen}},\ }\bibfield  {title} {\bibinfo {title} {{Ferguson distributions via P{\'o}lya urn schemes}},\ }\href@noop {} {\bibfield  {journal} {\bibinfo  {journal} {The Annals of Statistics}\ }\textbf {\bibinfo {volume} {1}},\ \bibinfo {pages} {353} (\bibinfo {year} {1973})}\BibitemShut {NoStop}%
\bibitem [{\citenamefont {Hill}\ \emph {et~al.}(1980)\citenamefont {Hill}, \citenamefont {Lane},\ and\ \citenamefont {Sudderth}}]{hill1980strong}%
  \BibitemOpen
  \bibfield  {author} {\bibinfo {author} {\bibfnamefont {B.~M.}\ \bibnamefont {Hill}}, \bibinfo {author} {\bibfnamefont {D.}~\bibnamefont {Lane}},\ and\ \bibinfo {author} {\bibfnamefont {W.}~\bibnamefont {Sudderth}},\ }\bibfield  {title} {\bibinfo {title} {A strong law for some generalized urn processes},\ }\href@noop {} {\bibfield  {journal} {\bibinfo  {journal} {The Annals of Probability}\ ,\ \bibinfo {pages} {214}} (\bibinfo {year} {1980})}\BibitemShut {NoStop}%
\bibitem [{\citenamefont {Muliere}\ \emph {et~al.}(2006)\citenamefont {Muliere}, \citenamefont {Paganoni},\ and\ \citenamefont {Secchi}}]{muliere2006randomly}%
  \BibitemOpen
  \bibfield  {author} {\bibinfo {author} {\bibfnamefont {P.}~\bibnamefont {Muliere}}, \bibinfo {author} {\bibfnamefont {A.~M.}\ \bibnamefont {Paganoni}},\ and\ \bibinfo {author} {\bibfnamefont {P.}~\bibnamefont {Secchi}},\ }\bibfield  {title} {\bibinfo {title} {A randomly reinforced urn},\ }\href@noop {} {\bibfield  {journal} {\bibinfo  {journal} {Journal of Statistical Planning and Inference}\ }\textbf {\bibinfo {volume} {136}},\ \bibinfo {pages} {1853} (\bibinfo {year} {2006})}\BibitemShut {NoStop}%
\bibitem [{\citenamefont {Fortini}\ \emph {et~al.}(2021)\citenamefont {Fortini}, \citenamefont {Petrone},\ and\ \citenamefont {Sariev}}]{fortini2021predictive}%
  \BibitemOpen
  \bibfield  {author} {\bibinfo {author} {\bibfnamefont {S.}~\bibnamefont {Fortini}}, \bibinfo {author} {\bibfnamefont {S.}~\bibnamefont {Petrone}},\ and\ \bibinfo {author} {\bibfnamefont {H.}~\bibnamefont {Sariev}},\ }\bibfield  {title} {\bibinfo {title} {{Predictive constructions based on measure-valued P{\'o}lya urn processes}},\ }\href@noop {} {\bibfield  {journal} {\bibinfo  {journal} {Mathematics}\ }\textbf {\bibinfo {volume} {9}},\ \bibinfo {pages} {2845} (\bibinfo {year} {2021})}\BibitemShut {NoStop}%
\bibitem [{\citenamefont {Sariev}\ \emph {et~al.}(2023)\citenamefont {Sariev}, \citenamefont {Fortini},\ and\ \citenamefont {Petrone}}]{sariev2023infinite}%
  \BibitemOpen
  \bibfield  {author} {\bibinfo {author} {\bibfnamefont {H.}~\bibnamefont {Sariev}}, \bibinfo {author} {\bibfnamefont {S.}~\bibnamefont {Fortini}},\ and\ \bibinfo {author} {\bibfnamefont {S.}~\bibnamefont {Petrone}},\ }\bibfield  {title} {\bibinfo {title} {Infinite-color randomly reinforced urns with dominant colors},\ }\href@noop {} {\bibfield  {journal} {\bibinfo  {journal} {Bernoulli}\ }\textbf {\bibinfo {volume} {29}},\ \bibinfo {pages} {132} (\bibinfo {year} {2023})}\BibitemShut {NoStop}%
\bibitem [{\citenamefont {Hoppe}(1984)}]{hoppe1984polya}%
  \BibitemOpen
  \bibfield  {author} {\bibinfo {author} {\bibfnamefont {F.~M.}\ \bibnamefont {Hoppe}},\ }\bibfield  {title} {\bibinfo {title} {{P{\'o}lya-like urns and the Ewens' sampling formula}},\ }\href@noop {} {\bibfield  {journal} {\bibinfo  {journal} {Journal of Mathematical Biology}\ }\textbf {\bibinfo {volume} {20}},\ \bibinfo {pages} {91} (\bibinfo {year} {1984})}\BibitemShut {NoStop}%
\bibitem [{\citenamefont {Aldous}(2006)}]{aldous2006exchangeability}%
  \BibitemOpen
  \bibfield  {author} {\bibinfo {author} {\bibfnamefont {D.~J.}\ \bibnamefont {Aldous}},\ }\bibfield  {title} {\bibinfo {title} {Exchangeability and related topics},\ }in\ \href@noop {} {\emph {\bibinfo {booktitle} {{\'E}cole d'{\'E}t{\'e} de Probabilit{\'e}s de Saint-Flour XIII—1983}}}\ (\bibinfo  {publisher} {Springer},\ \bibinfo {year} {2006})\ pp.\ \bibinfo {pages} {1--198}\BibitemShut {NoStop}%
\bibitem [{\citenamefont {Tria}\ \emph {et~al.}(2014)\citenamefont {Tria}, \citenamefont {Loreto}, \citenamefont {Servedio},\ and\ \citenamefont {Strogatz}}]{tria2014dynamics}%
  \BibitemOpen
  \bibfield  {author} {\bibinfo {author} {\bibfnamefont {F.}~\bibnamefont {Tria}}, \bibinfo {author} {\bibfnamefont {V.}~\bibnamefont {Loreto}}, \bibinfo {author} {\bibfnamefont {V.~D.~P.}\ \bibnamefont {Servedio}},\ and\ \bibinfo {author} {\bibfnamefont {S.~H.}\ \bibnamefont {Strogatz}},\ }\bibfield  {title} {\bibinfo {title} {The dynamics of correlated novelties},\ }\href@noop {} {\bibfield  {journal} {\bibinfo  {journal} {Scientific Reports}\ }\textbf {\bibinfo {volume} {4}},\ \bibinfo {pages} {5890} (\bibinfo {year} {2014})}\BibitemShut {NoStop}%
\bibitem [{\citenamefont {Merton}(1968)}]{merton1968matthew}%
  \BibitemOpen
  \bibfield  {author} {\bibinfo {author} {\bibfnamefont {R.~K.}\ \bibnamefont {Merton}},\ }\bibfield  {title} {\bibinfo {title} {{The Matthew effect in science: The reward and communication systems of science are considered}},\ }\href@noop {} {\bibfield  {journal} {\bibinfo  {journal} {Science}\ }\textbf {\bibinfo {volume} {159}},\ \bibinfo {pages} {56} (\bibinfo {year} {1968})}\BibitemShut {NoStop}%
\bibitem [{\citenamefont {Yule}(1925)}]{yule1925mathematical}%
  \BibitemOpen
  \bibfield  {author} {\bibinfo {author} {\bibfnamefont {G.~U.}\ \bibnamefont {Yule}},\ }\bibfield  {title} {\bibinfo {title} {{A mathematical theory of evolution, based on the conclusions of Dr. J. C. Willis}},\ }\href@noop {} {\bibfield  {journal} {\bibinfo  {journal} {Philosophical transactions of the Royal Society of London - B}\ }\textbf {\bibinfo {volume} {213}},\ \bibinfo {pages} {21} (\bibinfo {year} {1925})}\BibitemShut {NoStop}%
\bibitem [{\citenamefont {Zipf}(1949)}]{zipf1949human}%
  \BibitemOpen
  \bibfield  {author} {\bibinfo {author} {\bibfnamefont {G.~K.}\ \bibnamefont {Zipf}},\ }\href@noop {} {\emph {\bibinfo {title} {Human Behavior and the Principle of Least Effort: An Introduction to Human Ecology}}}\ (\bibinfo  {publisher} {Addison-Wesley Press},\ \bibinfo {address} {Cambridge, MA},\ \bibinfo {year} {1949})\BibitemShut {NoStop}%
\bibitem [{\citenamefont {{D. J. de Solla Price}}(1976)}]{price1976}%
  \BibitemOpen
  \bibfield  {author} {\bibinfo {author} {\bibnamefont {{D. J. de Solla Price}}},\ }\bibfield  {title} {\bibinfo {title} {A general theory of bibliometric and other cumulative advantage processes},\ }\href {https://doi.org/https://doi.org/10.1002/asi.4630270505} {\bibfield  {journal} {\bibinfo  {journal} {Journal of the American Society for Information Science}\ }\textbf {\bibinfo {volume} {27}},\ \bibinfo {pages} {292} (\bibinfo {year} {1976})}\BibitemShut {NoStop}%
\bibitem [{\citenamefont {Redner}(1998)}]{redner1998popular}%
  \BibitemOpen
  \bibfield  {author} {\bibinfo {author} {\bibfnamefont {S.}~\bibnamefont {Redner}},\ }\bibfield  {title} {\bibinfo {title} {How popular is your paper? an empirical study of the citation distribution},\ }\href@noop {} {\bibfield  {journal} {\bibinfo  {journal} {The European Physical Journal B-Condensed Matter and Complex Systems}\ }\textbf {\bibinfo {volume} {4}},\ \bibinfo {pages} {131} (\bibinfo {year} {1998})}\BibitemShut {NoStop}%
\bibitem [{\citenamefont {Gabaix}(1999)}]{gabaix1999zipf}%
  \BibitemOpen
  \bibfield  {author} {\bibinfo {author} {\bibfnamefont {X.}~\bibnamefont {Gabaix}},\ }\bibfield  {title} {\bibinfo {title} {Zipf's law and the growth of cities},\ }\href@noop {} {\bibfield  {journal} {\bibinfo  {journal} {American Economic Review}\ }\textbf {\bibinfo {volume} {89}},\ \bibinfo {pages} {129} (\bibinfo {year} {1999})}\BibitemShut {NoStop}%
\bibitem [{\citenamefont {Batty}(2008)}]{batty2008size}%
  \BibitemOpen
  \bibfield  {author} {\bibinfo {author} {\bibfnamefont {M.}~\bibnamefont {Batty}},\ }\bibfield  {title} {\bibinfo {title} {The size, scale, and shape of cities},\ }\href@noop {} {\bibfield  {journal} {\bibinfo  {journal} {Science}\ }\textbf {\bibinfo {volume} {319}},\ \bibinfo {pages} {769} (\bibinfo {year} {2008})}\BibitemShut {NoStop}%
\bibitem [{\citenamefont {{Tim Lewens}}(2018)}]{lewens2018}%
  \BibitemOpen
  \bibfield  {author} {\bibinfo {author} {\bibnamefont {{Tim Lewens}}},\ }\bibfield  {title} {\bibinfo {title} {{Stochasticity in cultural evolution: a revolution yet to happen}},\ }\href {https://doi.org/10.1007/s40656-017-0176-9} {\bibfield  {journal} {\bibinfo  {journal} {History and Philosophy of the Life Sciences}\ }\textbf {\bibinfo {volume} {40}},\ \bibinfo {pages} {9} (\bibinfo {year} {2018})}\BibitemShut {NoStop}%
\bibitem [{\citenamefont {Simon}(1955)}]{simon1955class}%
  \BibitemOpen
  \bibfield  {author} {\bibinfo {author} {\bibfnamefont {H.~A.}\ \bibnamefont {Simon}},\ }\bibfield  {title} {\bibinfo {title} {On a class of skew distribution functions},\ }\href@noop {} {\bibfield  {journal} {\bibinfo  {journal} {Biometrika}\ }\textbf {\bibinfo {volume} {42}},\ \bibinfo {pages} {425} (\bibinfo {year} {1955})}\BibitemShut {NoStop}%
\bibitem [{\citenamefont {Zanette}\ and\ \citenamefont {Montemurro}(2005)}]{Zanette2005}%
  \BibitemOpen
  \bibfield  {author} {\bibinfo {author} {\bibfnamefont {D.}~\bibnamefont {Zanette}}\ and\ \bibinfo {author} {\bibfnamefont {M.}~\bibnamefont {Montemurro}},\ }\bibfield  {title} {\bibinfo {title} {{Dynamics of Text Generation with Realistic Zipf's Distribution}},\ }\href@noop {} {\bibfield  {journal} {\bibinfo  {journal} {Journal of Quantitative Linguistics}\ }\textbf {\bibinfo {volume} {12}},\ \bibinfo {pages} {29} (\bibinfo {year} {2005})}\BibitemShut {NoStop}%
\bibitem [{\citenamefont {{P. Rosillo-Rodes, J. W. Zimmerman, L. Hébert-Dufresne and P. S. Dodds}}(2026)}]{rosillo_2026}%
  \BibitemOpen
  \bibfield  {author} {\bibinfo {author} {\bibnamefont {{P. Rosillo-Rodes, J. W. Zimmerman, L. Hébert-Dufresne and P. S. Dodds}}},\ }\href {https://doi.org/10.48550/arXiv.2604.13184} {\bibinfo {title} {{Simon's model does not produce Zipf's law: The fundamental rich-get-richer mechanism for any power-law size ranking}}} (\bibinfo {year} {2026}),\ \bibinfo {note} {arXiv preprint arXiv:2604.13184},\ \Eprint {https://arxiv.org/abs/2604.13184} {arXiv:2604.13184 [physics.soc-ph]} \BibitemShut {NoStop}%
\bibitem [{\citenamefont {Barab{\'a}si}\ and\ \citenamefont {Albert}(1999)}]{barabasi1999emergence}%
  \BibitemOpen
  \bibfield  {author} {\bibinfo {author} {\bibfnamefont {A.-L.}\ \bibnamefont {Barab{\'a}si}}\ and\ \bibinfo {author} {\bibfnamefont {R.}~\bibnamefont {Albert}},\ }\bibfield  {title} {\bibinfo {title} {Emergence of scaling in random networks},\ }\href@noop {} {\bibfield  {journal} {\bibinfo  {journal} {Science}\ }\textbf {\bibinfo {volume} {286}},\ \bibinfo {pages} {509} (\bibinfo {year} {1999})}\BibitemShut {NoStop}%
\bibitem [{\citenamefont {Gomez-Gardenes}\ and\ \citenamefont {Moreno}(2004)}]{gomez2004local}%
  \BibitemOpen
  \bibfield  {author} {\bibinfo {author} {\bibfnamefont {J.}~\bibnamefont {Gomez-Gardenes}}\ and\ \bibinfo {author} {\bibfnamefont {Y.}~\bibnamefont {Moreno}},\ }\bibfield  {title} {\bibinfo {title} {Local versus global knowledge in the barab{\'a}si-albert scale-free network model},\ }\href@noop {} {\bibfield  {journal} {\bibinfo  {journal} {Physical Review E}\ }\textbf {\bibinfo {volume} {69}},\ \bibinfo {pages} {037103} (\bibinfo {year} {2004})}\BibitemShut {NoStop}%
\bibitem [{\citenamefont {{Wikipedia contributors}}(2026{\natexlab{a}})}]{pharaons_eg_list}%
  \BibitemOpen
  \bibfield  {author} {\bibinfo {author} {\bibnamefont {{Wikipedia contributors}}},\ }\href {https://en.wikipedia.org/wiki/List_of_pharaohs} {\bibinfo {title} {List of pharaohs --- {Wikipedia}{,} the free encyclopedia}} (\bibinfo {year} {2026}{\natexlab{a}})\BibitemShut {NoStop}%
\bibitem [{\citenamefont {{Wikipedia contributors}}(2026{\natexlab{b}})}]{emperors_rom_list}%
  \BibitemOpen
  \bibfield  {author} {\bibinfo {author} {\bibnamefont {{Wikipedia contributors}}},\ }\href {https://en.wikipedia.org/wiki/List_of_Roman_emperors} {\bibinfo {title} {List of {Roman} emperors --- {Wikipedia}{,} the free encyclopedia}} (\bibinfo {year} {2026}{\natexlab{b}})\BibitemShut {NoStop}%
\bibitem [{\citenamefont {{Wikipedia contributors}}(2026{\natexlab{c}})}]{zars_ru_list}%
  \BibitemOpen
  \bibfield  {author} {\bibinfo {author} {\bibnamefont {{Wikipedia contributors}}},\ }\href {https://en.wikipedia.org/wiki/List_of_Russian_monarchs} {\bibinfo {title} {List of {Russian} monarchs --- {Wikipedia}{,} the free encyclopedia}} (\bibinfo {year} {2026}{\natexlab{c}})\BibitemShut {NoStop}%
\bibitem [{\citenamefont {{OrthodoxWiki contributors}}(2026)}]{patriarchs_const_list}%
  \BibitemOpen
  \bibfield  {author} {\bibinfo {author} {\bibnamefont {{OrthodoxWiki contributors}}},\ }\href {https://orthodoxwiki.org/List_of_Patriarchs_of_Constantinople} {\bibinfo {title} {List of {Patriarchs} of {Constantinople}}},\ \bibinfo {howpublished} {OrthodoxWiki} (\bibinfo {year} {2026})\BibitemShut {NoStop}%
\bibitem [{\citenamefont {{Wikipedia contributors}}(2026{\natexlab{d}})}]{kings_es_list}%
  \BibitemOpen
  \bibfield  {author} {\bibinfo {author} {\bibnamefont {{Wikipedia contributors}}},\ }\href {https://en.wikipedia.org/wiki/List_of_heads_of_state_of_Spain} {\bibinfo {title} {List of heads of state of {Spain} --- {Wikipedia}{,} the free encyclopedia}} (\bibinfo {year} {2026}{\natexlab{d}})\BibitemShut {NoStop}%
\bibitem [{\citenamefont {{Wikipedia contributors}}(2026{\natexlab{e}})}]{kings_dn_list}%
  \BibitemOpen
  \bibfield  {author} {\bibinfo {author} {\bibnamefont {{Wikipedia contributors}}},\ }\href {https://en.wikipedia.org/wiki/List_of_monarchs_of_Denmark} {\bibinfo {title} {List of monarchs of {Denmark} --- {Wikipedia}{,} the free encyclopedia}} (\bibinfo {year} {2026}{\natexlab{e}})\BibitemShut {NoStop}%
\bibitem [{\citenamefont {{The Editors of Encyclopaedia Britannica}}(2024)}]{kings_en_list}%
  \BibitemOpen
  \bibfield  {author} {\bibinfo {author} {\bibnamefont {{The Editors of Encyclopaedia Britannica}}},\ }\href {https://www.britannica.com/place/Kings-and-Queens-of-Britain-1856932} {\bibinfo {title} {Kings and {Queens} of {Britain}}},\ \bibinfo {howpublished} {Encyclopaedia Britannica} (\bibinfo {year} {2024})\BibitemShut {NoStop}%
\bibitem [{\citenamefont {Bianconi}\ and\ \citenamefont {Barab{\'a}si}(2001)}]{bianconi2001bose}%
  \BibitemOpen
  \bibfield  {author} {\bibinfo {author} {\bibfnamefont {G.}~\bibnamefont {Bianconi}}\ and\ \bibinfo {author} {\bibfnamefont {A.-L.}\ \bibnamefont {Barab{\'a}si}},\ }\bibfield  {title} {\bibinfo {title} {{Bose-Einstein condensation in complex networks}},\ }\href@noop {} {\bibfield  {journal} {\bibinfo  {journal} {Physical Review Letters}\ }\textbf {\bibinfo {volume} {86}},\ \bibinfo {pages} {5632} (\bibinfo {year} {2001})}\BibitemShut {NoStop}%
\bibitem [{\citenamefont {H{\'e}bert-Dufresne}\ \emph {et~al.}(2025{\natexlab{b}})\citenamefont {H{\'e}bert-Dufresne}, \citenamefont {Lovato}, \citenamefont {Burgio}, \citenamefont {Gleeson}, \citenamefont {Redner},\ and\ \citenamefont {Krapivsky}}]{hebert2025self}%
  \BibitemOpen
  \bibfield  {author} {\bibinfo {author} {\bibfnamefont {L.}~\bibnamefont {H{\'e}bert-Dufresne}}, \bibinfo {author} {\bibfnamefont {J.}~\bibnamefont {Lovato}}, \bibinfo {author} {\bibfnamefont {G.}~\bibnamefont {Burgio}}, \bibinfo {author} {\bibfnamefont {J.~P.}\ \bibnamefont {Gleeson}}, \bibinfo {author} {\bibfnamefont {S.}~\bibnamefont {Redner}},\ and\ \bibinfo {author} {\bibfnamefont {P.~L.}\ \bibnamefont {Krapivsky}},\ }\bibfield  {title} {\bibinfo {title} {Self-reinforcing cascades: A spreading model for beliefs or products of varying intensity or quality},\ }\href@noop {} {\bibfield  {journal} {\bibinfo  {journal} {Physical Review Letters}\ }\textbf {\bibinfo {volume} {135}},\ \bibinfo {pages} {087401} (\bibinfo {year} {2025}{\natexlab{b}})}\BibitemShut {NoStop}%
\bibitem [{\citenamefont {{O. Kolodny, N. Creanza and M. W. Feldman}}(2016)}]{kolodny2016}%
  \BibitemOpen
  \bibfield  {author} {\bibinfo {author} {\bibnamefont {{O. Kolodny, N. Creanza and M. W. Feldman}}},\ }\bibfield  {title} {\bibinfo {title} {{Game-Changing Innovations: How Culture Can Change the Parameters of Its Own Evolution and Induce Abrupt Cultural Shifts}},\ }\href {https://doi.org/10.1371/journal.pcbi.1005302} {\bibfield  {journal} {\bibinfo  {journal} {PLOS Computational Biology}\ }\textbf {\bibinfo {volume} {12}},\ \bibinfo {pages} {e1005302} (\bibinfo {year} {2016})}\BibitemShut {NoStop}%
\bibitem [{\citenamefont {{D. Centola, J. Becker, D. Brackbill and A. Baronchelli}}(2018)}]{centola2018}%
  \BibitemOpen
  \bibfield  {author} {\bibinfo {author} {\bibnamefont {{D. Centola, J. Becker, D. Brackbill and A. Baronchelli}}},\ }\bibfield  {title} {\bibinfo {title} {{Experimental evidence for tipping points in social convention}},\ }\href {https://doi.org/10.1126/science.aas8827} {\bibfield  {journal} {\bibinfo  {journal} {Science}\ }\textbf {\bibinfo {volume} {360}},\ \bibinfo {pages} {1116} (\bibinfo {year} {2018})}\BibitemShut {NoStop}%
\bibitem [{\citenamefont {{Shakshi Singhal and Adarsh Anand and Ompal Singh}}(2020)}]{singhal2020}%
  \BibitemOpen
  \bibfield  {author} {\bibinfo {author} {\bibnamefont {{Shakshi Singhal and Adarsh Anand and Ompal Singh}}},\ }\bibfield  {title} {\bibinfo {title} {{Studying dynamic market size-based adoption modeling \& product diffusion under stochastic environment}},\ }\href {https://doi.org/10.1016/j.techfore.2020.120285} {\bibfield  {journal} {\bibinfo  {journal} {Technological Forecasting \& Social Change}\ }\textbf {\bibinfo {volume} {161}},\ \bibinfo {pages} {120285} (\bibinfo {year} {2020})}\BibitemShut {NoStop}%
\bibitem [{\citenamefont {Raaijmakers}()}]{raaijmakers2024}%
  \BibitemOpen
  \bibfield  {author} {\bibinfo {author} {\bibfnamefont {J.}~\bibnamefont {Raaijmakers}},\ }\bibfield  {title} {\bibinfo {title} {The social meaning of names},\ }in\ \href {https://doi.org/10.1484/M.USML-EB.5.133524} {\emph {\bibinfo {booktitle} {Writing Names in Medieval Sacred Spaces}}},\ \bibinfo {series and number} {Utrecht Studies in Medieval Literacy},\ pp.\ \bibinfo {pages} {299--307}\BibitemShut {NoStop}%
\end{thebibliography}%

\clearpage

\newpage

\section{Methods}

\subsection{Name standardization}

We standardize the name data by isolating the primary component of authority in each dynastic tradition. In the case of the Roman emperors, we isolate the \textit{cognomen}. In the case of monarchs, we restrict the analysis to the primary regnal proper name (e.g., King George VI is considered as George). We limit the analysis of papal names to the chosen proper name (e.g., Jorge Bergoglio is Francis). In the case of the pharaohs, we discard the first four titular names, isolating the \textit{nomen}, as it is the case of the Ottoman sultans, where we analyze the core regnal name, omitting all introductory and concluding titular formulas.

\subsection{Prestige-driven innovation-reinforcement model}

We model the temporal evolution of papal names as a stochastic process combining innovation and prestige-driven reinforcement. At each election time $t$, the set of names that have already appeared is denoted by $A(t)$, with $M(t)=|A(t)|$. Each active name $i\in A(t)$ is characterized by its abundance $n_i(t)$, defined as the number of times the name has been used, and by its accumulated prestige $x_i(t)$. The total prestige in the system is
\begin{equation}
X(t)=\sum_{j\in A(t)} x_j(t).\label{eq:X_t}
\end{equation}

At election $t$, a new name is introduced with probability $\rho(t)$. With complementary probability $1-\rho(t)$, an already active name is reused. Conditional on reuse, name $i$ is selected with probability proportional to its accumulated prestige. Therefore, the unconditional probability that an already active name $i\in A(t)$ is chosen is
\begin{equation}
p_i(t)=\left[1-\rho(t)\right]\frac{x_i(t)}{X(t)}.
\end{equation}
The probability that the election introduces a new name is
\begin{equation}
p_{\mathrm{new}}(t)=\rho(t).
\end{equation}

Whenever a name is selected, its prestige increases by a random increment $K$, independently drawn from a prestige-increment distribution $P(K)$, with mean $\mu=\mathbb{E}[K]$ and variance $\sigma^2=\mathrm{Var}(K)$. If an already active name $i$ is reused, then
\begin{equation}
n_i(t+1)=n_i(t)+1,
\qquad
x_i(t+1)=x_i(t)+K(t).
\end{equation}
If innovation occurs, a new name $m\notin A(t)$ enters the system with
\begin{equation}
n_m(t+1)=1,
\qquad
x_m(t+1)=K(t).
\end{equation}
Thus, innovation does not create empty types: each new name enters with one observed use and an initial prestige sampled from the same distribution that governs later prestige increments.

The total number of name elections is
\begin{equation}
N(t)=\sum_{i\in A(t)} n_i(t),
\end{equation}
and grows as $N(t)=N_0+t$. The expected number of distinct names follows
\begin{equation}
M(t)=M_0+\int_0^t \rho(s)\,ds. \label{eq:M_t}
\end{equation}
\medskip

\subsection{Prestige distribution}

In the RIP model, each reign contributes a positive prestige increment $K$ to the name under which it occurs. We model these increments as independent random variables drawn from a truncated power-law distribution,
\begin{equation}
P(K)=C_{\alpha,\kappa}K^{-\alpha},
\qquad
1\leq K\leq \kappa ,
\label{eq:prestige_distribution}
\end{equation}
where $\alpha>0$ is the power-law exponent, $\kappa\geq1$ is the upper cutoff, and $C_{\alpha,\kappa}$ is the normalization constant. The exponent $\alpha$ controls the heaviness of the tail, whereas $\kappa$ fixes the largest possible prestige increment. Thus, smaller values of $\alpha$ or larger values of $\kappa$ both increase the potential role of rare but highly prestigious reigns.
\medskip


We can connect the parameters of the truncated power law distribution with the mean and variance as follows. The mean prestige increment is
\begin{equation}
\mu(\alpha,\kappa)
=
\mathbb{E}[K]
=
\frac{1-\alpha}{2-\alpha}
\frac{\kappa^{2-\alpha}-1}{\kappa^{1-\alpha}-1},
\label{eq:prestige_mean}
\end{equation}
and the variance of prestige increments 
\begin{equation}
\sigma^2(\alpha,\kappa)
=
\frac{1-\alpha}{3-\alpha}
\frac{\kappa^{3-\alpha}-1}{\kappa^{1-\alpha}-1}
-
\left[
\frac{1-\alpha}{2-\alpha}
\frac{\kappa^{2-\alpha}-1}{\kappa^{1-\alpha}-1}
\right]^2 .
\label{eq:prestige_variance_explicit}
\end{equation}
Note that in the special cases $\alpha=1,2,3$ the corresponding expressions can be obtained as logarithmic limits.

\subsubsection{Fitting the prestige distribution}

Because each naming tradition requires different magnitudes of rare events, the prestige distribution must be tailored to each dynastic time series. To achieve this, we evaluate an $(\alpha, \kappa)$ parameter grid across 100 simulations, identifying the optimal $(\alpha, \kappa)$ point that minimizes the RMSE between the average simulated variance and the empirical variance. For the innovation rate, we compute a smoothed $\rho(t)$ using a centered rolling window of length 3. The results of this optimization process are presented in Table~\ref{tab:gridsearch}.

\begin{table}[htbp]
\centering
\caption{Optimal fit parameters for $P(K)$ in Eq.~\eqref{eq:prestige_distribution} obtained through RMSE minimization.}
\label{tab:gridsearch}
\begin{tabular}{lcc}
\toprule
\textbf{Tradition} & $\kappa$ & $\alpha$ \\
\midrule
Spanish monarchs & 105 & 4.98 \\\hline
Danish monarchs & 374 & 3.90 \\\hline
English monarchs & 70 & 1.82 \\\hline
Russian tsars  & 1 & - \\\hline
Roman emperors & 585 & 2.71 \\\hline
Pharaohs & 244 & 1.31 \\\hline
Holy Roman emperors & 1 & - \\\hline
Constantinople patriarchs & 330 & 4.22 \\\hline
Ottoman sultans & 1 & - \\\hline
Popes & 209 & 3.07 \\
\bottomrule
\end{tabular}
\end{table}

\subsection{Quantifying the empirical variance}

To quantify the heterogeneity of name abundances, we consider the empirical variance over the active repertoire,
\begin{equation}
\mathcal{\overline{V}}_n(t)
=
\frac{\overline{Q}(t)}{M(t)}
-
\left[
\frac{N(t)}{M(t)}
\right]^2, \label{eq:variance}
\end{equation} where we have introduced the second abundance moment
\begin{equation}
\overline{Q}(t)=\sum_{i\in A(t)} n_i^2(t).
\end{equation}

We approximate the expected evolution of this quantity through a continuous-time second-moment closure. We now aim to compute the theoretical variance
\begin{equation}
\overline{\mathcal{V}}_n(t)
=
\frac{\overline{Q}(t)}{M(t)}
-
\left[
\frac{N(t)}{M(t)}
\right]^2. \label{eq:variance_theo}
\end{equation}

Besides $\overline{Q}(t)$, we define the abundance--prestige mixed moment
\begin{equation}
\overline{Y}(t)=\sum_{i\in A(t)} n_i(t)x_i(t),
\end{equation}
and the prestige second moment
\begin{equation}
\overline{W}(t)=\sum_{i\in A(t)} x_i^2(t).
\end{equation}
As derived in the Supplementary Note 2, for an arbitrary innovation function $\rho(t)$, the formal solution in Eq. (\ref{eq:variance}) can be written explicitly by computing Eq. (\ref{eq:M_t}) and the second moments as follows:
\begin{eqnarray}
   \overline{W}(t)
&=&
e^{2G_1(t)}
\left[
W_0
+
(\mu^2+\sigma^2)
\int_0^t e^{-2G_1(s)}\,\mathrm{d}s
\right], \\
\overline{Y}(t)
&=&
e^{G_1(t)}\\&&
\left[
Y_0
+
\int_0^t
e^{-G_1(s)}
\left(
\mu
+
\left[1-\rho(s)\right]
\frac{\overline{W}(s)}{\mu N(s)}
\right)
\mathrm{d}s
\right],\nonumber\\
\overline{Q}(t)
&=&
Q_0+t
+
\frac{2}{\mu}
\int_0^t
\left[1-\rho(s)\right]
\frac{\overline{Y}(s)}{N(s)}
\,\mathrm{d}s,\label{eq:Q_t}
\end{eqnarray}
being $G_1(t)
=
\int_0^t
\frac{1-\rho(u)}{N(u)}\,\mathrm{d}u$.

\subsection{RIP vs. RGR in the limit scenario of no innovation}

To get an analytical insight into the excess variance induced by the heterogeneous prestige distribution, let us consider the closed-repertoire limit, \(\rho(t)=0\), for which the number of active names remains fixed, \(M(t)=M_0\), while the total number of choices grows as \(N(t)=N_0+t\). 


For the prestige-driven model, the empirical abundance variance scales as (see the full derivation in Supplementary Note~2)
\begin{eqnarray}
 &&\overline{\mathcal V}_{n,\mathrm{RIP}}(t)\label{eq:v_prestige_noinnovation}
\\&&\sim
\left[
\frac{1}{M_0}
\left(
\frac{W_0}{\mu^2 N_0^2}
+
\frac{1}{N_0}
\left[
1+\frac{\sigma^2}{\mu^2}
\right]
\right)
-
\frac{1}{M_0^2}
\right]
(N_0+t)^2,\nonumber
\end{eqnarray}
where \(W_0=\sum_{i=1}^{M_0}x_i^2(0)\), \(\mu=\mathbb{E}[K]\), and \(\sigma^2=\mathrm{Var}(K)\). For the pure preferential-attachment model, in which reused names are selected proportionally to their current abundance, the corresponding leading-order expression is
\begin{equation}
\overline{\mathcal V}_{n,\mathrm{RGR}}(t)
\sim
\left[
\frac{1}{M_0}
\left(
\frac{Q_0}{N_0^2}
+
\frac{1}{N_0}
\right)
-
\frac{1}{M_0^2}
\right]
(N_0+t)^2,
\label{eq:v_RGR_noinnovation}
\end{equation}
with \(Q_0=\sum_{i=1}^{M_0}n_i^2(0)\). Thus, in the absence of innovation, both reinforcement mechanisms generate quadratic growth of the abundance variance, but heterogeneous prestige increments increase the prefactor through the dimensionless ratio \(\sigma^2/\mu^2\).

If both processes are initialized consistently, \(x_i(0)=\mu n_i(0)\), so that \(W_0=\mu^2 Q_0\), Eqs.~(\ref{eq:v_prestige_noinnovation})--(\ref{eq:v_RGR_noinnovation}) yield Eq. (\ref{eq:comparison}).

\renewcommand{\figurename}{Supplementary Fig.}
\renewcommand{\tablename}{Supplementary Table}
\renewcommand{\theequation}{S.\arabic{equation}}

\setcounter{equation}{0}
\setcounter{figure}{0}

\onecolumngrid
\newpage

\section{Supplementary Note 1: Onomastics}

\subsection{The role of prestige in each dataset}

Name selection varies significantly across ruling traditions. In several monarchies, rulers traditionally reigned under their primary baptismal name. In other royal systems monarchs selected from a composite of multiple baptismal names, as exemplified by King George VI of England, whose full baptismal name was Albert Frederick Arthur George. In contrast, ancient Egyptian pharaohs systematically adopted regnal names upon ascension. In the case of popes, name change did not establish itself as an institutional trend until approximately the year 900 with the election of John X (with only two prior pontiffs changing their names).

Due to these operational differences, the symbolic prestige of names acts differently in each ruling system. We operationalize prestige as the probability of a name being selected based on the historical legacy and achievements of its prior holders. Within dynastic, hereditary frameworks, the assignment of a given name to an heir reflects a naming tradition in a given family or lineage. However, when an emperor or pope consciously adopts a specific name upon installation, it serves as an exercise in political marketing, drawing upon the accumulated prestige of that lineage to project institutional continuity, stability, and power. 
\medskip

\begin{figure*}[t]
  \centering
  \includegraphics[width=\linewidth]{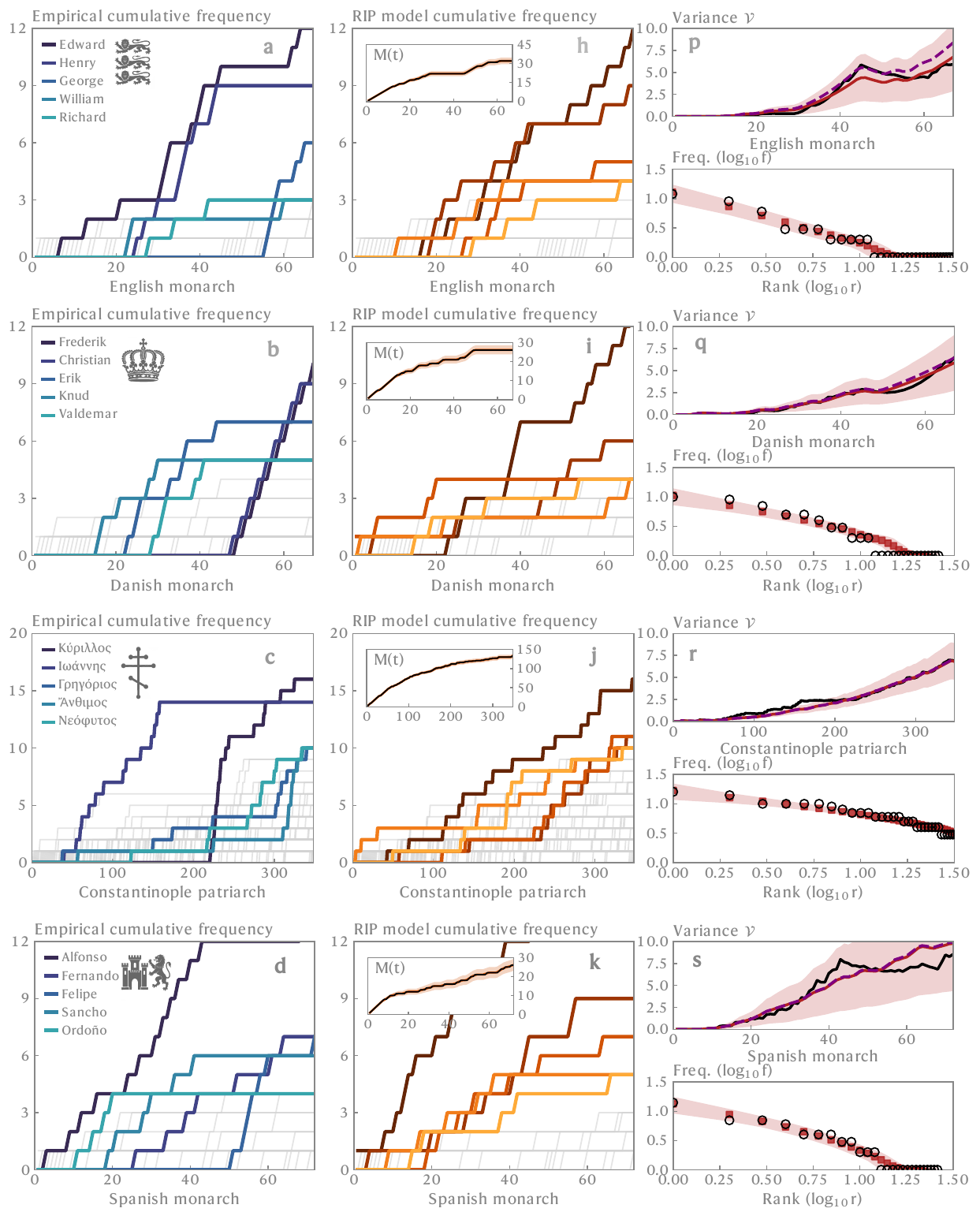}
  \label{fig:s1_1}
\end{figure*}

\begin{figure*}[t]
  \centering
  \includegraphics[width=\linewidth]{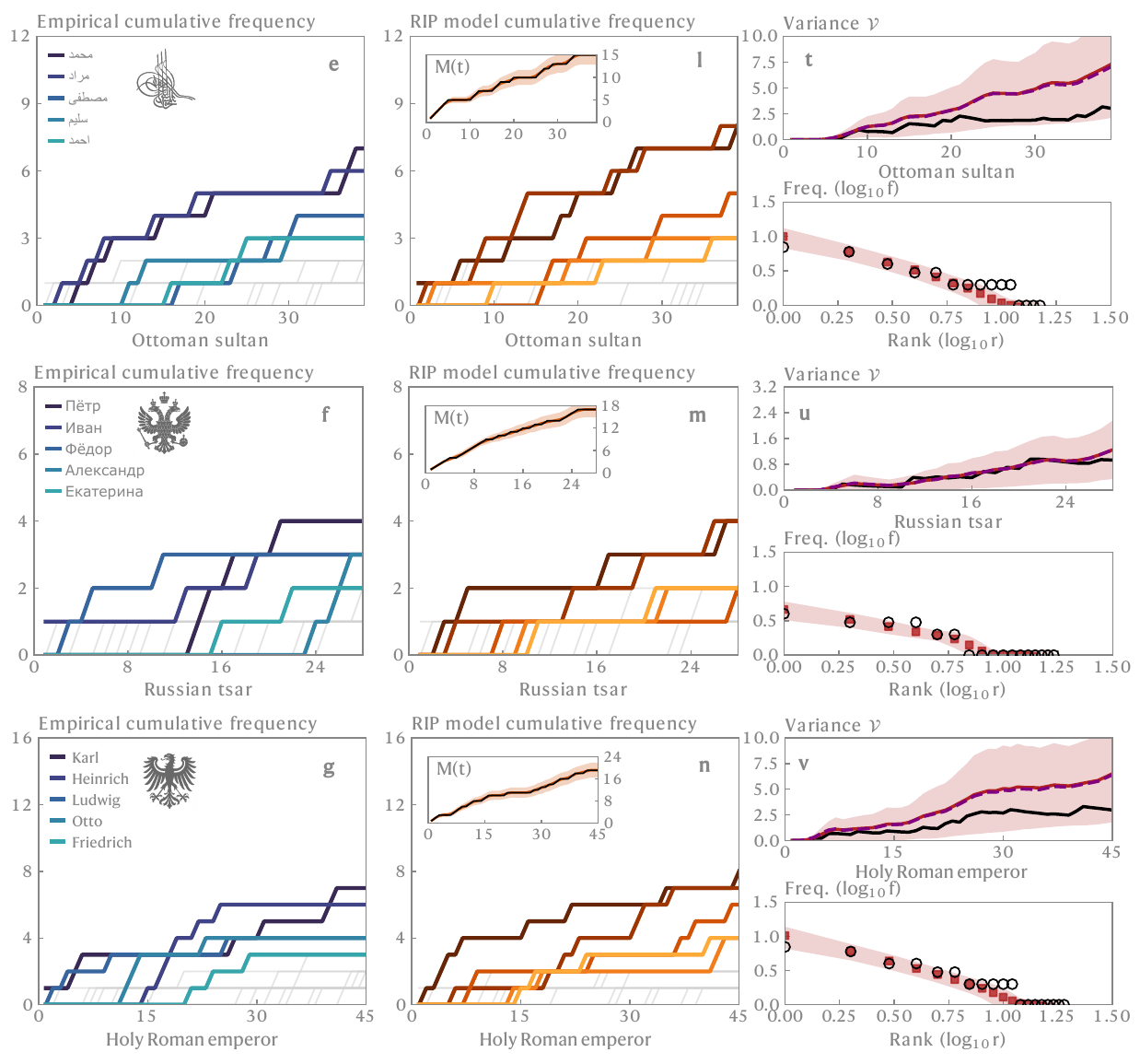}
\caption{\justifying\textbf{a}-\textbf{g} Number of Danish monarchs, English monarchs, patriarchs of Constantinople, Spanish monarchs, SRGE Holy Roman emperors, Ottoman sultans and Russian tsars, respectively, bearing each name across chronologically ordered reigns. \textbf{h}-\textbf{n} Output of a single run of the RIP model.  The insets show the evolution of the number of distinct names $(M(t)$ in bottom rows) over time.
\textbf{p}-\textbf{v} Top: Variance of name frequencies in the data compared with 1000 simulations of the RIP model and with the analytical expression from Eqs.~(15)-(20). Bottom: rank-frequency distribution of names. The plots show the frequency $f_i$ of each name $i$ as a function of its rank $r_i$, comparing the empirical data with the mean outcome of 1000 simulations of the RIP model $\pm1\sigma$.}
  \label{fig:s1_2}
\end{figure*}

\subsection{Papal onomastics}

The bishops of Rome have occupied a special place in the naming culture over the last two millennia. In the Middle Ages, a name was never a neutral label. Naming was a social act that inscribed an individual into kinship networks, local traditions, and patterns of patronage, and it was often a religious act that invoked the protection or imitation of a saint~\cite{raaijmakers2024}. By the seventh century in Western Europe, local customs largely determined naming patterns, yet a gradual convergence toward a more universal repertoire can be detected, even as communities clung to their own onomastic habits. At the same time, demographic growth and new administrative demands drove a shift from single names to hereditary surnames, while the stock of given names shrank in what has been described as an onomastic deflation~\cite{raaijmakers2024}. Before Francis, the papal names had been repeated for more than a millennium.
\medskip

The papacy’s own naming history divides neatly into two long stages. Across roughly the first millennium, almost every pope retained the name he bore at the moment of election~\cite{nau1993}. Romans, Greeks, Africans, Jews, and others ascended the Holy See and brought with them names rooted in different linguistic and cultural worlds. 
The early papal onomasticon thus mirrors the diversity and relative fluidity of late antique and early medieval Christian society.
\medskip

The transition from this first stage of stability to a second one governed by papal self-naming was triggered by an intricate political history. Between the mid-ninth and mid-eleventh centuries, Rome and its church were buffeted by internal factionalism, intervention from Frankish and German emperors, competition with Constantinople, and threats from Muslim powers~\cite{devinne2006}. In this context, an “anomalous” interlocking series of Popes John and Benedict emerged between the ninth and eleventh centuries, during which the two names reached exceptionally high rates of reuse~\cite{devinne2006}. Their recurrence tracked the influence of powerful Roman lineages such as the Crescentii and the counts of Tusculum and showed how strongly emperors were steering papal affairs at different moments~\cite{devinne2006,poole1917}. Several of these Johns and Benedicts were murdered, deposed, accused of simony or immorality, or installed and removed by secular patrons. 
\medskip

It is in the wake of this crisis that papal onomastic innovation abruptly disappears, starting with John X in 914. Gradually, papal self-naming becomes established. What had been an occasional expedient (e.g., Mercurius taking the name John II in 533 to avoid a pagan theonym) turned into a custom endowed with theological and political meaning. The reforming German popes installed by Henry III in the mid-eleventh century chose new names that linked them to ancient martyrs and revered bishops of Rome even as they signaled a break with the scandalous recent past~\cite{devinne2006}. By changing their names, popes could also imitate the biblical pattern of Peter and Paul, dramatize the spiritual transformation implied in their election, and underline the separation of their office from secular family ties and worldly careers, in harmony with the broader reform agenda that sought clerical celibacy and autonomy~\cite{nau1993}.
\medskip

Once introduced, the new practice quickly crystallized into an onomastic tradition with its own internal logic. Name choice allowed a newly elected pope to weave himself into a chain of predecessors and saints while projecting a future program for his pontificate~\cite{devinne2006}; now, popes are telling stories when choosing a name. High medieval popes overwhelmingly selected names of early bishops of Rome. In later centuries, the pattern shifted from emulating the distant patristic past to honoring more recent predecessors, family connections, religious orders, or the symbolic content of a name. In all these cases, the underlying criterion remained the same: names carried accumulated prestige, and choosing one meant tapping into the symbolic capital built by earlier pontiffs or revered figures. For example, Pius II famously drew on Virgil’s epithet “pius Aeneas” when selecting his own name; John XXIII resurrected a neglected, even tarnished, name for reasons ranging from family homage to local patronage and modest expectations of a short pontificate; and John Paul I crafted an unprecedented double name to signal continuity with both John XXIII’s pastoral warmth and Paul VI’s administrative skill~\cite{devinne2006,nau1993}. More recent popes, such as Benedict XVI, have explicitly explained their choices as bids for peace, reform, or alignment with particular spiritual traditions, making the act of naming itself a first narrative gesture of their pontificates~\cite{devinne2006}.
\medskip

The long pattern of reusing established papal names ended with the election of Pope Francis in 2013. He was the first Jesuit pope, the first from Latin America,  and the first pope named Francis. His choice introduced a new phase in papal naming, one that has continued with the recent election of Pope Leo XIV.

\subsection{Onomastics of powerful dynasties}

In Supplementary Fig. \ref{fig:s1_2} we expand the analysis of Fig. 1 to the remaining dynasties.

\clearpage
\newpage

\section{Supplementary Note 2: BIC and AIC for assessing the goodness of the fit}

The Akaike Information Criterion (AIC) and the Bayesian Information Criterion (BIC) are statistical tools used for model selection that balance how well a model fits the data with its complexity to prevent overfitting. AIC focuses primarily on minimizing information loss and optimizing predictive performance on new datasets, which can sometimes lead it to choose overly complex models in large samples. Conversely, BIC introduces a stricter penalty that depends on the total sample size, favoring more parsimonious, simpler models and demonstrating statistical consistency by asymptotically selecting the true data-generating model if it is among the choices.

When we only have RMSE data available instead of a direct log-likelihood value, AIC and BIC can still be computed under the assumption that the model's residuals are independent and identically normally distributed. In this case, the maximized log-likelihood simplifies into a direct function of the sample size and the log-RMSE. We have then $$\text{AIC} = 2n\ln(\text{RMSE}) + 2k$$ and $$\text{BIC} = 2n\ln(\text{RMSE}) + k\ln(n),$$ where $n$ represents the number of observations and $k$ denotes the total number of parameters. Therefore, the lowest the BIC or AIC, the better. 

In Supplementary Fig.~\ref{fig:s3} we show the performance of RIP, rich-get-richer and random models for each dataset.

\begin{figure*}[h]
  \centering
  \includegraphics[width=\linewidth]{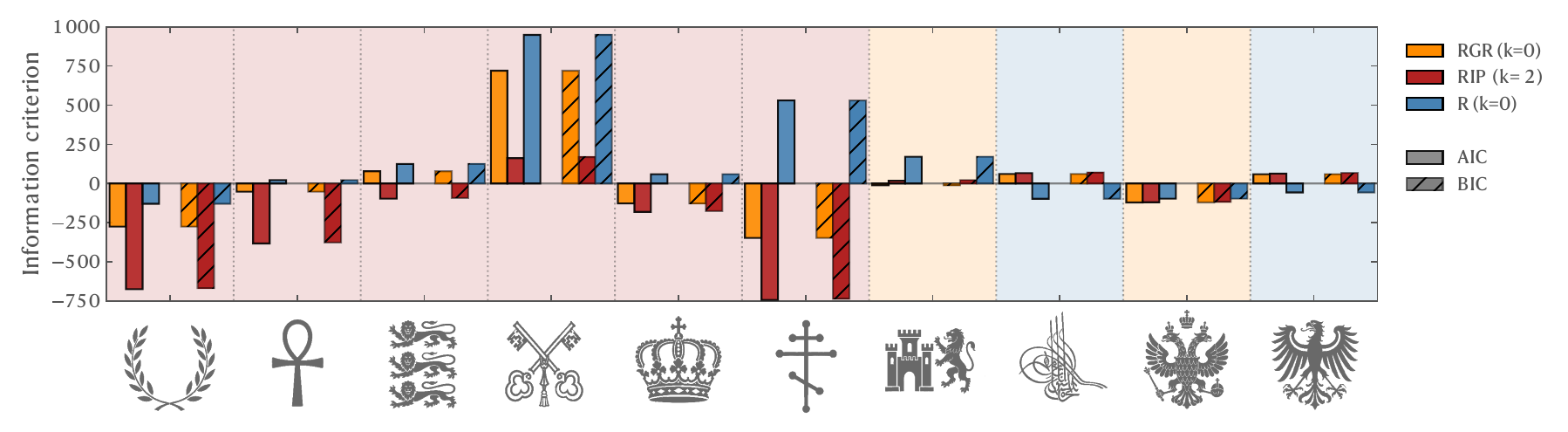}
\caption{\justifying\textbf{Goodness of the fit for prestige, rich-get-richer and random models} in terms of two additional metrics, the Bayesian and the Akaike Information Criterions. The horizontal labels correspond to the symbols shown in Fig.~2.}
  \label{fig:s3}
\end{figure*}

\clearpage

\newpage

\section{Supplementary Note 3: The Reinforcement
and Innovation through Prestige (RIP) model}

\subsection*{Reinforcement
and Innovation through Prestige (RIP) model}

We consider a stochastic model for the temporal evolution of names. At each name election, either a new name is introduced or one of the names that have already appeared in the historical record is reused.
\medskip

Let $A(t)$ denote the set of active names before election $t$, and let $M(t)=|A(t)|$ be the number of distinct names observed up to that point. Each active name $i\in A(t)$ is characterized by two variables: its abundance $n_i(t)$, defined as the number of times the name has been used, and its accumulated prestige $x_i(t)$. The total prestige in the system is

\begin{equation}
X(t)=\sum_{j\in A(t)}x_j(t).
\end{equation}

At election $t$, innovation occurs with probability $\rho(t)\in[0,1]$.
If innovation occurs, a new name $m\notin A(t)$ is introduced. If innovation does not occur, one of the already active names is selected with probability proportional to its accumulated prestige. Therefore, for an already active name $i\in A(t)$, the unconditional probability of being chosen at election $t$ is

\begin{equation}
p_i(t)
=
[1-\rho(t)]\frac{x_i(t)}{X(t)}.
\label{eq:selection_probability}
\end{equation}

The probability that election $t$ introduces a new name is

\begin{equation}
p_{\mathrm{new}}(t)=\rho(t).
\end{equation}

We define the innovation indicator $J_t$ which takes value $J_t=1$ if election introduces a new name and $J_t=0$ otherwise, with $\mathbb{E}[J_t]=\rho(t)$.
We also define the reinforcement indicator
$Z_t=1-J_t$, so that $Z_t=1$ when election $t$ reuses an already existing name, and $Z_t=0$ when election $t$ introduces a new name.

Whenever a name is selected, its prestige increases by a random increment $K$, independently drawn from a prestige-increment distribution $P(K)$. We denote its mean, variance and second moment by $\mu=\mathbb{E}[K]>0$,  $\sigma^2=\mathrm{Var}(K)$ and $m_2=\mathbb{E}[K^2]=\mu^2+\sigma^2$.

Therefore, if an already active name $i\in A(t)$ is selected at election $t$, then its abundance and prestige evolve as

\begin{eqnarray}
n_i(t+1)&=&n_i(t)+1,\\
x_i(t+1)&=&x_i(t)+K(t),
\end{eqnarray}
where $K(t)$ is drawn from $P(K)$. All other active names $j\neq i$ remain unchanged during that election.

If innovation occurs, a new name $m\notin A(t)$ enters the system. Its abundance and prestige are initialized as

\begin{eqnarray}
    n_m(t+1)&=&1,\\
    x_m(t+1)&=&K(t),
    \label{eq:innovation_initial_prestige}
\end{eqnarray}
where, again,  $K(t)$ is drawn from $P(K)$. Thus, innovation does not introduce an empty type: a new name enters the system with one observed use and with an initial prestige drawn from the same prestige-increment distribution that governs reinforcement events.
After innovation, the active set is updated as $A(t+1)=A(t)\cup\{m\}$.

The model therefore combines two mechanisms. The function $\rho(t)$ controls the opening of the name repertoire through the introduction of new names. The prestige variables $x_i(t)$ control the reinforcement dynamics among already active names. Large values of $\rho(t)$ favor innovation and repertoire expansion, whereas small values of $\rho(t)$ favor the reuse of established names through prestige-driven reinforcement.

\subsection{Growth of the repertoire and variance of name abundances}

We now derive the expected growth of the number of distinct names and the time evolution of the variance of name abundances in the RIP model.
Let
\begin{equation}
N(t)=\sum_{i\in A(t)} n_i(t)
\end{equation}

be the total number of name elections recorded up to time $t$. Since exactly one name is assigned at each election,

\begin{equation}
N(t)=N_0+t,
\label{eq:N_t}
\end{equation}

where $N_0$ is the number of elections already present in the initial condition.

The number of active names evolves according to

\begin{equation}
M(t)=M_0+\sum_{s=0}^{t-1}J_s.
\end{equation}

Since $\mathbb{E}[J_s]=\rho(s)$,
we obtain $
\mathbb{E}[M(t)]
=
M_0+\sum_{s=0}^{t-1}\rho(s).
$
In a continuous-time approximation, this becomes
\begin{equation}
M(t)
=
M_0+\int_0^t \rho(s)\,\mathrm{d}s.
\label{eq:M_general}
\end{equation}

We next consider the variance of the abundance distribution over the set of active names. We define the empirical variance

\begin{equation}
\mathcal V(t)
=
\frac{1}{M(t)}
\sum_{i\in A(t)}
\left[
n_i(t)-\frac{N(t)}{M(t)}
\right]^2.
\end{equation}

Introducing the second moment

\begin{equation}
Q(t)=\sum_{i\in A(t)}n_i^2(t),
\end{equation}

the variance can be written as

\begin{equation}
\mathcal V(t)
=
\frac{\overline Q(t)}{M(t)}
-
\left[
\frac{N(t)}{M(t)}
\right]^2.
\label{eq:variance_from_Q}
\end{equation}

Therefore, the problem reduces to obtaining the evolution of $Q(t)$.

\subsubsection{Determining the evolution of $Q(t)$}

If election $t$ is an innovation event, a new name enters with abundance $n_m(t+1)=1$. Hence,

\begin{equation}
\Delta Q(t)|_{\text{innovation}}=1.
\end{equation}

If election $t$ is a reinforcement event and an already active name $i$ is selected, then

\begin{equation}
n_i(t+1)=n_i(t)+1,
\end{equation}

and therefore

\begin{equation}
\Delta Q(t)|_{\text{reinforcement}}
=
[n_i(t)+1]^2-n_i^2(t)
=
2n_i(t)+1.
\end{equation}

Using the prestige-driven selection probability in Eq. (\ref{eq:selection_probability}), we obtain the exact conditional evolution given all the previous information about the state of the system $F_t$,

\begin{equation}
\mathbb{E}[\Delta Q(t)\mid \mathcal F_t]
=
\rho(t)
+
[1-\rho(t)]
\sum_{i\in A(t)}
\frac{x_i(t)}{X(t)}
[2n_i(t)+1].
\label{eq:Delta_Q}
\end{equation}

We can introduce the abundance-prestige second moment 
\begin{equation}
Y(t)=\sum_{i\in A(t)}n_i(t)x_i(t),
\end{equation}
and since $
\sum_{i\in A(t)}x_i(t)/X(t)
=1$, Eq. (\ref{eq:Delta_Q}) simplifies to

\begin{equation}
\mathbb{E}[\Delta Q(t)\mid \mathcal F_t]
=
1
+
2[1-\rho(t)]
\frac{Y(t)}{X(t)}.
\label{eq:Q_exact}
\end{equation}

Equation~\eqref{eq:Q_exact} is exact, but it is not closed because the second moment of abundances depends on the evolution of the prestige and the correlation between abundance and prestige.

To close the calculation, we derive the evolution of $X(t)$ and $Y(t)$. The total prestige increases by one prestige increment at every election, either because a new name enters with initial prestige $K$, or because an already active name receives an additional prestige increment. Thus,

\begin{equation}
\mathbb{E}[\Delta X(t)\mid \mathcal F_t]=\mu,
\end{equation}

and, at the level of expectations, $\overline X(t)\simeq X_0+\mu t$. If the initial prestige is proportional to the initial number of elections, this can be written as

\begin{equation}
\overline X(t)\simeq \mu N(t).
\end{equation}

Then, lets derive $Y(t)$. If innovation occurs, a new name enters with $n_m(t+1)=1$ and $x_m(t+1)=K(t)$,
and hence

\begin{equation}
\Delta Y(t)|_{\text{innovation}}=K(t).
\end{equation}

If an already active name $i$ is selected, then

\begin{equation}
n_i(t+1)=n_i(t)+1,
\qquad
x_i(t+1)=x_i(t)+K(t).
\end{equation}

Therefore, the contribution of name $i$ to $Y(t)$ changes as

\begin{align}
\Delta Y(t)|_{\text{reinforcement}}
&=
[n_i(t)+1][x_i(t)+K(t)]
-
n_i(t)x_i(t)
\\
&=
x_i(t)+K(t)n_i(t)+K(t).
\end{align}

Averaging over the prestige increment and over the selected name gives

\begin{align}
\mathbb{E}[\Delta Y(t)\mid \mathcal F_t]
&=
\rho(t)\mu
+
[1-\rho(t)]
\sum_{i\in A(t)}
\frac{x_i(t)}{X(t)}
\left[
x_i(t)+\mu n_i(t)+\mu
\right]
\\
&=\mu
+
[1-\rho(t)]
\left[
\frac{W(t)}{X(t)}
+
\mu\frac{Y(t)}{X(t)}
\right], 
\label{eq:Delta-Y}
\end{align}
where we have introduced the second moment of prestige,

\begin{equation}
W(t)=\sum_{i\in A(t)}x_i^2(t).
\end{equation}

Again, Eq. (\ref{eq:Delta-Y}) is not closed, and depends on the evolution of $W(t)$. If innovation occurs, a new name with prestige $K$ is introduced, and therefore

\begin{equation}
\Delta W(t)=K^2.
\end{equation}

If an already active name $i$ is selected, its prestige changes as

\begin{equation}
x_i(t+1)=x_i(t)+K(t),
\end{equation}

so that

\begin{equation}
\Delta W(t)
=
[x_i(t)+K(t)]^2-x_i^2(t)
=
2K(t)x_i(t)+K^2(t).
\end{equation}

Conditioned on a reinforcement event, the probability that the selected
active name is $i$ is $x_i(t)/X(t)$. Equivalently, the unconditional probability of selecting name $i$ is
$[1-\rho(t)]x_i(t)/X(t)$, consistently with Eq.~(S.2).
Averaging over the prestige increment and over the prestige-driven choice of the selected name gives

\begin{eqnarray}
\mathbb{E}[\Delta W(t)\mid \mathcal F_t]
&=&
\rho(t)m_2
+
[1-\rho(t)]
\sum_{i\in A(t)}
\frac{x_i(t)}{X(t)}
\left[
2\mu x_i(t)+m_2
\right]\\
&=&
m_2
+
2\mu[1-\rho(t)]
\frac{W(t)}{X(t)},
\end{eqnarray}

where we have used again that the prestige-weighted probabilities sum to one. Note that the variance of prestige increments does not enter directly in the update of the abundances $n_i(t)$. Instead, it modifies the dispersion of accumulated prestige through $W(t)$. This affects the abundance--prestige coupling $Y(t)$, which in turn controls the growth of $Q(t)$ and therefore the variance of name abundances.
\medskip

In a continuous-time second-moment closure, we replace random quantities by their expected values and recall that $X(t)\simeq \mu N(t)$. We finally obtain the closed system

\begin{eqnarray}
\frac{\mathrm{d}M}{\mathrm{d}t}\label{eq:closed_M}
&=&
\rho(t),\\
\frac{\mathrm{d}\overline W}{\mathrm{d}t}
&=&
\mu^2+\sigma^2
+
2(1-\rho(t))\frac{\overline W(t)}{N(t)},\label{eq:closed_W}\\
\frac{\mathrm{d}\overline Y}{\mathrm{d}t}
&=&
\mu
+
(1-\rho(t))
\left[
\frac{\overline W(t)}{\mu N(t)}
+
\frac{\overline Y(t)}{N(t)}
\right],\label{eq:closed_Y}\\
\frac{\mathrm{d}\overline Q}{\mathrm{d}t}
&=&
1
+
2(1-\rho(t))
\frac{\overline Y(t)}{\mu N(t)}.
\label{eq:closed_Q}
\end{eqnarray}

The expected empirical variance is then approximated by

\begin{equation}
\overline{\mathcal V}(t)
\simeq
\frac{\overline Q(t)}{M(t)}
-
\left[
\frac{N(t)}{M(t)}
\right]^2.
\label{eq:closed_variance}
\end{equation}

This provides the desired general expression for arbitrary innovation function $\rho(t)$. Equations~\eqref{eq:closed_M}--\eqref{eq:closed_Q} can be integrated numerically for any prescribed $\rho(t)$, and Eq.~\eqref{eq:closed_variance} gives the corresponding predicted variance of the abundance distribution.

For completeness, the formal solution for arbitrary $\rho(t)$ can also be written explicitly. Let

\begin{equation}
G_1(t)=\int_0^t \frac{1-\rho(u)}{N(u)}\,\mathrm{d}u,
\qquad
G_2(t)=2G_1(t).
\end{equation}

Then

\begin{eqnarray}
M(t) &=& M_0+\int_0^t \rho(s)\,\mathrm{d}s, \\
\overline W(t) &=& e^{G_2(t)} \left[ W_0 + (\mu^2+\sigma^2) \int_0^t e^{-G_2(s)}\,\mathrm{d}s \right], \\
\overline Y(t) &=& e^{G_1(t)} \left[ Y_0 + \int_0^t e^{-G_1(s)} \left( \mu + (1-\rho(s))\frac{\overline W(s)}{\mu N(s)} \right) \mathrm{d}s \right], \\
\overline Q(t) &=& Q_0+t + \frac{2}{\mu} \int_0^t (1-\rho(s)) \frac{\overline Y(s)}{N(s)} \,\mathrm{d}s.
\end{eqnarray}

Substituting these expressions into Eq.~\eqref{eq:closed_variance} gives the abundance variance for a general innovation function $\rho(t)$.

\subsection{Limit scenario of no innovation}

In the non-innovation case, $\rho(t)=0$, the number of names is fixed, and therefore the evolution of the active names reads $M(t)=M_0$.
We use Eq. (\ref{eq:N_t}), recalling that $\rho(t)=0$. Since $N(t)=N_0+t$, we have $dN/\mathrm{d}t=1$, and we can equivalently use $N$ as the time variable. Therefore, Eq. (\ref{eq:closed_W}) can be solved explicitly as

\begin{eqnarray}
\overline W(t)
&=&
N^2(t)
\left[
\frac{W_0}{N_0^2}
+
(\mu^2+\sigma^2)
\left(
\frac{1}{N_0}
-
\frac{1}{N(t)}
\right)
\right]\\
&=& B_W N^2(t) - (\mu^2+\sigma^2)N(t),\label{eq:W_t}
\end{eqnarray}
being
\begin{equation}
B_W = \frac{W_0}{N_0^2} + \frac{\mu^2+\sigma^2}{N_0}.\label{eq:Bw}
\end{equation}
Hence, the leading-order behavior is
\begin{equation}
\overline W(t) \sim B_W N^2(t),
\end{equation}
and shows explicitly that the heterogeneity of the prestige increments modifies the leading growth of the accumulated-prestige second moment. We now propagate this leading contribution through the moment hierarchy.  
Again, we can equivalently use $N$ as the time variable. This way, Eq. (\ref{eq:closed_Y}) becomes

\begin{equation}
\frac{\mathrm{d}\overline Y}{dN} - \frac{\overline Y}{N} = \mu + \frac{\overline W(N)}{\mu N}.
\end{equation}
Using Eq. (\ref{eq:W_t}) and integrating, we obtain
\begin{equation}
\overline Y(t) = N(t) \left\{ \frac{Y_0}{N_0} + \frac{B_W}{\mu} \left[ N(t)-N_0 \right] - \frac{\sigma^2}{\mu} \log\left[ \frac{N(t)}{N_0} \right] \right\}.
\end{equation}
The dominant term is thus
\begin{equation}
\overline Y(t) \sim \frac{B_W}{\mu}N^2(t).\label{eq:Y_t}
\end{equation}
We now use the leading contribution in Eq. (\ref{eq:Y_t}) to solve the differential equation for the abundance second moment, Eq. (\ref{eq:closed_Q}). After retainig the dominant linear contribution, we obtain 
\begin{equation}
\overline Q(t) \sim \frac{B_W}{\mu^2}N^2(t).\label{eq:Q_t}
\end{equation}
Finally, since the repertoire is closed, $M(t)=M_0$, and using Eqs. (\ref{eq:N_t})-(\ref{eq:Bw})-(\ref{eq:Q_t}) the approximated empirical abundance variance in Eq. (\ref{eq:closed_variance}) is
\begin{equation}
\overline{\mathcal V}(t) \sim \left[ \frac{1}{M_0} \left( \frac{W_0}{\mu^2N_0^2} + \frac{1}{N_0} \left[ 1+\frac{\sigma^2}{\mu^2} \right] \right) - \frac{1}{M_0^2} \right] (N_0+t)^2, \label{eq:v_pre}
\end{equation}
which for large $t$ yields a quadratic asymptotic behavior $\overline{\mathcal V}(t)
\sim\,t^2$. Importantly,  this equation shows explicitly how the prestige-increment variance enters the predicted abundance variance. The quadratic scaling comes from the absence of innovation, whereas the prefactor is increased by the dimensionless prestige heterogeneity $\sigma^2/\mu^2$.

\subsection*{Limit scenario of full innovation}

If every election introduces a new name, then there are no reinforcement events. If the initial condition consists only of names that have appeared once, then $N_0=M_0$ and $Q_0=M_0$. Consequently,

\begin{equation}
N(t)=M_0+t,
\qquad
M(t)=M_0+t,
\qquad
Q(t)=M_0+t,
\end{equation}
and therefore when we substitute in Eq. (\ref{eq:closed_variance}) we obtain that 
\begin{equation}
\mathcal V(t)=0.
\end{equation}
In this full-innovation limit, the prestige distribution still affects the accumulated prestige variables $x_i(t)$, but it has no effect on the abundance variance because prestige never enters the selection of an already active name.

\subsection{Limit scenario of constant innovation in the large-times limit}

For constant innovation probability, $\rho(t)=\rho$, the not only the number of elections grows asymptotically as $N(t)\sim t$, but also the number of active names. Integrating Eq. (\ref{eq:M_general}), we find that $M(t)\sim \rho t$.

Since both $N(t)$ and $M(t)$ grow linearly with time, $Q(t)$ must also scale with $t$ in order that the variance in Eq. (\ref{eq:closed_variance}) approaches a finite asymptotic value and does not diverge. This implies that $\mathrm{d}\overline{Q}/\mathrm{d}t$ must approach a constant. 

We introduce the rescaled moments
\begin{equation}
    U(t)=\frac{\overline{W}(t)}{\mu^2}, \qquad V(t)=\frac{\overline{Y}(t)}{\mu}.
\end{equation}
For constant $\rho$, their equations are
\begin{eqnarray}
  \frac{\mathrm{d}U}{\mathrm{d}t} &=& 1+\frac{\sigma^2}{\mu^2} + 2(1-\rho)\frac{U}{N},\label{eq:rescaled_u}\\  
  \frac{\mathrm{d}V}{\mathrm{d}t} &=& 1 + (1-\rho)\left[ \frac{U}{N} + \frac{V}{N} \right],\label{eq:rescaled_v}
\end{eqnarray}
so that Eq. (\ref{eq:closed_variance}) becomes
\begin{equation}
    \frac{\mathrm{d}\overline{Q}}{\mathrm{d}t} = 1 + 2(1-\rho)\frac{V}{N}.\label{eq:rescaled_q}
\end{equation}

Again using $dN/\mathrm{d}t=1$, we look for linear asymptotic scalings $U(N)\sim c_U N$, $V(N)\sim c_V N$, $\overline{Q}(N)\sim c_Q N$.

Substituting $U(N)\sim c_U N$ into Eq. (\ref{eq:rescaled_u}) yields
\begin{equation}
    c_U = \frac{1+\sigma^2/\mu^2}{2\rho-1}.
\end{equation}

Next, substituting $U(N)\sim c_U N$ and $V(N)\sim c_V N$ into Eq. (\ref{eq:rescaled_v}), we get
\begin{equation}
c_V = \frac{1}{\rho} \left[ 1+ (1-\rho)\frac{1+\sigma^2/\mu^2}{2\rho-1} \right],
\end{equation}
and using Eq. (\ref{eq:rescaled_q}),
\begin{equation}
  c_Q = 1 + \frac{2(1-\rho)}{\rho} \left[ 1+ (1-\rho)\frac{1+\sigma^2/\mu^2}{2\rho-1} \right].  
\end{equation}
Substituting all the rescaling in Eq. (\ref{eq:closed_variance}), we find that the asymptotic abundance variance is
\begin{eqnarray}
    \overline{\mathcal V}(\infty) &=& \frac{c_Q}{\rho} - \frac{1}{\rho^2} \\
&=& \frac{1-\rho}{\rho^2} \left[ 1+ \frac{2(1-\rho)}{2\rho-1} \left( 1+\frac{\sigma^2}{\mu^2} \right) \right].
\end{eqnarray}

The denominator $2\rho-1$ shows that this finite asymptotic expression is well defined only for $\rho>1/2$. For $\rho\leq\rho_c= 1/2$, reinforcement dominates strongly enough that the second moments no longer grow linearly with $N(t)$; consequently, $\overline Q(t)/M(t)$ diverges and the abundance variance does not approach a finite large-time limit.

\begin{figure*}[h]
  \centering
  \includegraphics[width=\linewidth]{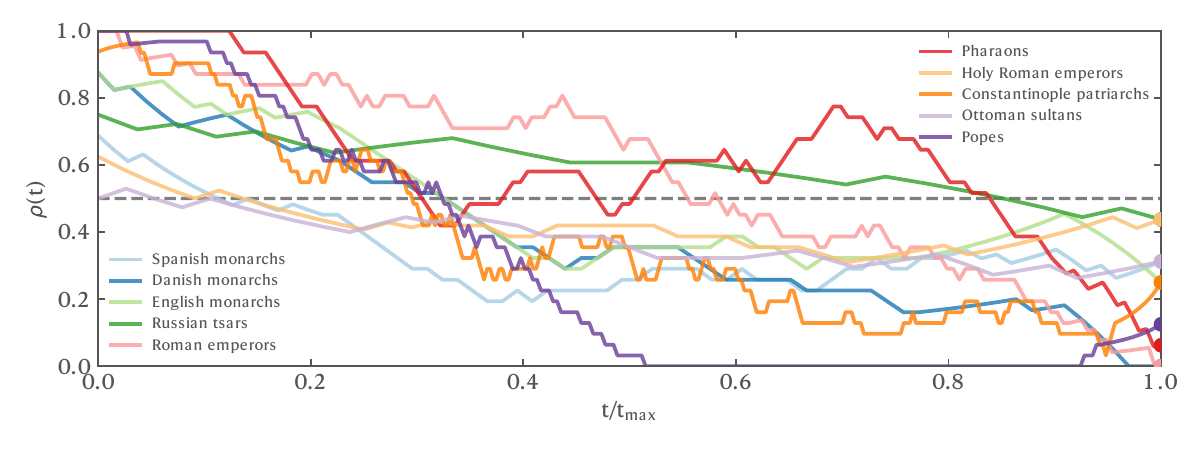}
\caption{\justifying \textbf{Empirical innovation profiles across dynastic naming traditions.} 
Time-dependent innovation rate $\rho(t)$ for the ten historical datasets, plotted as a function of normalized time $t/t_{\max}$. The innovation rate is estimated from the empirical sequence of name choices using a centered rolling window of length 30 for visual clarity.}
  \label{fig:s2}
\end{figure*}

Supplementary Fig. \ref{fig:s2} shows that the empirical innovation rates are not constant across the historical sequences. Instead, after an initial innovation-dominated regime, $\rho(t)$ decreases in all datasets, reflecting the progressive exhaustion or closure of the available name repertoire. Nonetheless, this decay places the long-time dynamics in an effective low-innovation regime (below the aforementioned $\rho_c=1/2$), where reinforcement is no longer diluted by the arrival of new names. Therefore, the continued growth of the variance trajectories observed in Fig.~1g--i and Supplementary Fig.~\ref{fig:s1_2} is the expected outcome of the empirical innovation profiles.

\subsection{Comparison with rich-get-richer and random sampling}

We now compare the abundance variance generated by the prestige-driven model with that generated by pure rich-get-richer.
In the pure rich-get-richer model, established names are selected with probability

\begin{equation}
p_i^{\mathrm{RGR}}(t)
=
[1-\rho(t)]\frac{n_i(t)}{N(t)},
\end{equation}
which does not depend on the prestige. As a consequence, the second moment of the abundance distribution now obeys the closed equation
\begin{equation}
\frac{\mathrm{d}\overline Q_{\mathrm{RGR}}}{\mathrm{d}t}
=
1
+
2(1-\rho(t))\frac{\overline Q_{\mathrm{RGR}}(t)}{N(t)},\label{eq:Q_PA}
\end{equation}

Note that pure rich-get-richer is recovered when prestige is exactly proportional to abundance, $x_i(t)=\mu n_i(t)$. In that case, the rescaled moments $U(t)=V(t)=\overline Q_{\mathrm{RGR}}(t)$, 
and the prestige-driven equation reduces to the rich-get-richer equation.

To compare the two models, assume the same initial condition and take $x_i(0)=\mu n_i(0)$, so that $U(0)=V(0)=Q_{\mathrm{RGR}}(0)$. We define the differences $\Delta U(t)=U(t)-\overline Q_{\mathrm{RGR}}(t)$, $\Delta V(t)=V(t)-\overline Q_{\mathrm{RGR}}(t)$, and $\Delta Q(t)=\overline Q_{\mathrm{RIP}}(t)-\overline Q_{\mathrm{RGR}}(t)$.
Subtracting the rich-get-richer equation from the prestige-driven closure yields

\begin{eqnarray}
  \frac{\mathrm{d}\Delta U}{\mathrm{d}t}
&=&
\frac{\sigma^2}{\mu^2}
+
2\left(1-\rho(t)\right)\frac{\Delta U(t)}{N(t)},\\  
\frac{\mathrm{d}\Delta V}{\mathrm{d}t}
&=&
\left(1-\rho(t)\right)
\frac{\Delta U(t)+\Delta V(t)}{N(t)},\\
\frac{\mathrm{d}\Delta Q}{\mathrm{d}t}
&=&
2\left(1-\rho(t)\right)\frac{\Delta V(t)}{N(t)}.
\end{eqnarray}

Since $1-\rho(t)\ge 0$, $N(t)>0$, $\sigma^2/\mu^2\ge 0$, and the initial differences vanish, these equations imply that $\Delta Q(t)\ge 0$. Therefore, $\overline Q_{\mathrm{RIP}}(t)
\ge
\overline Q_{\mathrm{RGR}}(t)$.
Since both models have the same $N(t)$ and the same expected number of names $M(t)$, the difference in abundance variance is

\begin{equation}
\overline{\mathcal V}_{\mathrm{RIP}}(t)
-
\overline{\mathcal V}_{\mathrm{RGR}}(t)
\simeq
\frac{
\overline Q_{\mathrm{RIP}}(t)
-
\overline Q_{\mathrm{RGR}}(t)
}
{M(t)}.
\end{equation}

Hence,

\begin{equation}
\overline{\mathcal V}_{\mathrm{RIP}}(t)
\ge
\overline{\mathcal V}_{\mathrm{RGR}}(t).
\end{equation}

The inequality becomes an equality if prestige increments are deterministic, $\sigma^2=0$, provided that $x_i(0)=\mu n_i(0)$. In that case, prestige remains proportional to abundance and the prestige-driven model collapses exactly onto pure rich-get-richer.
\medskip

Finally, we compare with a null model in  which name reuse is not preferential. Innovation is kept unchanged: with probability $\rho(t)$, a new name enters with abundance one. However, when no innovation occurs, an already active name is selected uniformly at random.
Thus, established names are selected with probability 
\begin{equation}
p_i^{\mathrm{R}}(t)
=
[1-\rho(t)]\frac{1}{M(t)},
\end{equation}
which does not depend on the prestige nor in the abundances. Therefore, now the second moment of the abundance distribution obeys the closed equation
\begin{equation}
\frac{\mathrm{d}\overline Q_{\mathrm{R}}}{\mathrm{d}t}
=
1
+
2(1-\rho(t))\frac{N(t)}{M(t)}.\label{eq:Rand}
\end{equation}

This should be compared with pure rich-get-richer in Eq. (\ref{eq:Q_PA}). The difference between the two mechanisms is clear. Random reuse is controlled only by the mean abundance per active name, $N(t)/M(t)$, whereas rich-get-richer is controlled by the abundance-weighted mean, $\overline Q_{\mathrm{RGR}}(t)/N(t)$. Since the abundance variance is non-negative,
\begin{equation}
\frac{\overline Q_{\mathrm{RGR}}(t)}{M(t)}
-
\left[
\frac{N(t)}{M(t)}
\right]^2
\geq 0,
\end{equation}
we have $\overline Q_{\mathrm{RGR}}(t) / N(t) \geq N(t) / M(t)$. Therefore, $\mathrm{d}\overline Q_{\mathrm{RGR}} / \mathrm{d}t \geq \mathrm{d}\overline Q_{\mathrm{R}} / \mathrm{d}t$.
With the same initial condition, this implies $\overline Q_{\mathrm{RGR}}(t)
\geq
\overline Q_{\mathrm{R}}(t)$.
Since both models have the same $N(t)$ and the same expected number of active names $M(t)$, the corresponding abundance variances satisfy
\begin{equation}
\overline{\mathcal V}_{\mathrm{RGR}}(t)
\geq
\overline{\mathcal V}_{\mathrm{R}}(t).
\end{equation}

Combining this inequality with the comparison between prestige-driven reinforcement and pure rich-get-richer, we obtain the hierarchy
\begin{equation}
\overline{\mathcal V}_{\mathrm{RIP}}(t)
\geq
\overline{\mathcal V}_{\mathrm{RGR}}(t)
\geq
\overline{\mathcal V}_{\mathrm{R}}(t).
\end{equation}

The first inequality follows from the additional heterogeneity induced by random prestige increments, whereas the second follows from the reinforcement of already abundant names in rich-get-richer. Random reuse lacks both amplification mechanisms and therefore produces the smallest abundance variance within this comparison.

\subsubsection{Asymptotic Limits in the Closed-Repertoire Regime (No Innovation)}

We consider again the closed-repertoire limit where the discovery of novel elements is suppressed, locking the total number of unique names at a fixed baseline, i.e. $M(t)=M_0$.
\medskip

\noindent\textbf{RGR model:} In the baseline pure rich-get-richer model, provided $\rho(t)=0$, Eq. (\ref{eq:Q_PA}) can be solved explicitly using the integrating factor as
\begin{equation}
\overline Q_{\mathrm{RGR}}(t) = \left[ \frac{Q_0}{N_0^2} + \frac{1}{N_0} - \frac{1}{N(t)} \right]N^2(t).
\end{equation}

As chronological time advances to the asymptotic limit ($t \to \infty$, implying $N(t) \to \infty$), the transient $1/N(t)$ term vanishes, leaving the leading-order scaling behavior:
\begin{equation}
\overline Q_{\mathrm{RGR}}(t) \sim \left[ \frac{Q_0}{N_0^2} + \frac{1}{N_0} \right] N^2(t).
\end{equation}

Then, substituting in Eq. (\ref{eq:closed_variance}) and recalling that $M(t)=M_0$ and $N(t)=N_0+t$ gives:
\begin{equation}
\overline{\mathcal{V}}_{\mathrm{RGR}}(t) \sim \left[\frac{1}{M_0}\left( \frac{Q_0}{N_0^2} + \frac{1}{N_0} \right) - \frac{1}{M_0^2}\right](N_0+t)^2. \label{eq:v_pa}
\end{equation}
\medskip

\noindent\textbf{Random model:} In the random reuse model, names are selected uniformly from the closed pool without historical feedback. Provided $\rho(t)=0$, Eq. (\ref{eq:Rand}) can be solved explicitly as 
\begin{equation}
\overline Q_{\mathrm{R}}(t) = Q_0 + [N(t)-N_0] + \frac{N^2(t)-N_0^2}{M_0}.
\end{equation}

Then, substituting in Eq. (\ref{eq:closed_variance}) and recalling that $M(t)=M_0$ and $N(t) - N_0 = t$ gives
\begin{eqnarray}
\overline{\mathcal{V}}_{\mathrm{R}}(t) &=& \left( \frac{Q_0}{M_0} - \frac{N_0^2}{M_0^2} \right) + \frac{N(t)-N_0}{M_0}\\
&=& \mathcal{V}(0) + \frac{t}{M_0}
\end{eqnarray}
Substituting $N(t) - N_0 = t$ simplifies the continuous-closure random reuse variance to a strictly linear relationship
\begin{equation}
\overline{\mathcal{V}}_{\mathrm{R}}(t) = \mathcal{V}(0) + \frac{t}{M_0}.\label{eq:v_rand}
\end{equation}

The analytical expressions derived within this continuous-time closure framework in Eqs. (\ref{eq:v_pre})-(\ref{eq:v_pa})-(\ref{eq:v_rand}) confirm the macroscopic asymptotic hierarchy
\begin{equation}
\overline{\mathcal{V}}_{\mathrm{R}}(t)=\mathcal{O}(t), \qquad \overline{\mathcal{V}}_{\mathrm{RGR}}(t)=\mathcal{O}(t^2), \qquad \overline{\mathcal{V}}_{\mathrm{RIP}}(t)=\mathcal{O}(t^2).
\end{equation}

Clearly, the random model has a lower variance than the other two, as the growth depends linearly on $t$. Moreover, we can substract the two asymptotic variances given by Eqs. (\ref{eq:v_pre})-(\ref{eq:v_pa}) to isolate the excess heterogeneity generated by our model
\begin{equation}
\overline{\mathcal{V}}_{\mathrm{RIP}}(t) - \overline{\mathcal{V}}_{\mathrm{RGR}}(t) \sim \frac{1}{M_0N_0} \frac{\sigma^2}{\mu^2} N^2(t).
\end{equation}

This formula demonstrates that while both mechanisms drive a quadratic accumulation of variance, prestige-driven reinforcement yields a strictly larger prefactor whenever the internal weights are updated via heterogeneous increments ($\sigma^2 > 0$).
\medskip

We note that solving the process using a continuous-moment closure introduces minor finite-size discretization differences relative to an exact discrete-time calculation. For example, in the pure rich-get-richer model, the true discrete-time P\'{o}lya urn prefactor replaces the continuous-closure term:
\begin{equation}
\left[ \frac{Q_0+N_0}{N_0^2} \right] \quad \longrightarrow \quad \left[ \frac{Q_0+N_0}{N_0(N_0+1)} \right].
\end{equation}

Similarly, for the random reuse model, the exact discrete-time calculation yields a linear slope of $(M_0-1)/M_0^2$ instead of the continuous-closure slope of $1/M_0$. These variations are strictly finite-size discretization corrections that vanish as $N_0 \to \infty$; they leave the underlying scaling laws and the asymptotic hierarchy qualitatively unchanged.

\subsubsection{Limit scenario of full innovation}

If every election introduces a new name, then there are no reinforcement events. If the initial condition consists only of names that have appeared once, then $N_0=M_0$ and $Q_0=M_0$ regardless of the model. Consequently, in all cases
\begin{equation}
\overline{\mathcal V}(t)=0.
\end{equation}

\subsubsection{Limit scenario of constant innovation in the large-times limit}

We now consider $\rho(t)=\rho$ in the $t\rightarrow\infty$ limit, where we recall that as $N(t) \sim t$ and $M(t) \sim \rho t$, $N(t)/M(t) \to 1/\rho$. 
\medskip

\noindent\textbf{RGR Model:}  For the macroscopic collective abundance variance to converge toward a stable, non-divergent finite ceiling at long times, the second moment $\overline{Q}_{\mathrm{RGR}}(N)$ in Eq. (\ref{eq:Q_PA}) must scale linearly with the total system size $N$ as
$\overline{Q}_{\mathrm{RGR}}(N) \sim c_{\mathrm{RGR}} N$. Substituting the ansatz in  Eq. (\ref{eq:Q_PA}) we are able to obtain that
\begin{eqnarray}
\overline{\mathcal{V}}_{\mathrm{RGR}}(\infty) &=& \frac{c_{\mathrm{RGR}}}{\rho} - \frac{1}{\rho^2}\\
&=&  \frac{1 - \rho}{\rho^2(2\rho - 1)}.
\end{eqnarray}

As in the prestige model, when the entry rate of novel elements drops below this critical threshold ($\rho \leq 1/2$), the ongoing influx of innovation is no longer sufficient to dilute the competitive compounding advantages of early-mover elements. In this weak-innovation regime, the raw second moment $\overline{Q}_{\mathrm{RGR}}(t)$ grows strictly faster than linearly with respect to $N(t)$ (scaling as $\mathcal{O}(t \ln t)$ at $\rho = 1/2$ and as $\mathcal{O}(t^{2-2\rho})$ for $\rho < 1/2$). As a result, the collective abundance variance diverges indefinitely, preventing the system from ever settling into a stable, stationary structural equilibrium.
\medskip

\noindent\textbf{Random Model:} In the random case, we can directly substitute the stationary state limit $N(t)/M(t) \to 1/\rho$ directly into Eq. (\ref{eq:Rand}) and integrate the equation to obtain yields the leading order asymptotic rate of change:
\begin{equation}
\overline{Q}_{\mathrm{R}}(t) \sim  \frac{2-\rho}{\rho} t.
\end{equation}

Substituting in  Eq. (\ref{eq:Q_PA}) and taking the $t\rightarrow\infty$ we are able to obtain that
\begin{equation}
\overline{\mathcal{V}}_{\mathrm{R}}(\infty) = \frac{1-\rho}{\rho^2}.
\end{equation}

In this case, the steady-state solution is unconditionally stable and structurally bounded for any active innovation rate $\rho > 0$.

Synthesizing these results for the strong-innovation regime ($\rho > 1/2$) where all the variances are defined, we recover that
\begin{equation}
\overline{\mathcal{V}}_{\mathrm{RIP}}(\infty) \geq \overline{\mathcal{V}}_{\mathrm{RGR}}(\infty) \geq \overline{\mathcal{V}}_{\mathrm{R}}(\infty).
\end{equation}

The first inequality ($\overline{\mathcal{V}}_{\mathrm{RIP}}(\infty) \geq \overline{\mathcal{V}}_{\mathrm{RGR}}(\infty)$) isolates the additional macroscopic heterogeneity injected by the stochastic updates of prestige increments (quantified by the squared coefficient of variation $\sigma^2/\mu^2$). This inequality collapses to a perfect identity if and only if prestige increments are purely deterministic ($\sigma^2=0$). The second inequality ($\overline{\mathcal{V}}_{\mathrm{RGR}}(\infty) \geq \overline{\mathcal{V}}_{\mathrm{R}}(\infty)$) emerges because history-dependent empirical feedback loops actively amplify early random sampling fluctuations, whereas random reuse continually dampens fluctuations back toward a uniform baseline.

\clearpage
\newpage

\section{Supplementary Note 4: Isomorphic Network model}

The RIP model can be represented as a growing network whose connected
components correspond to names. This duality offers a topological interpretation of naming history.

\subsection{The network formulation}

Each node represents one ruler, pontificate, or
reign, and is endowed at birth with an intrinsic prestige value
$\eta_v=K_v$, drawn independently from the prestige-increment distribution
$P(K)$. At time $t$, let $C_i(t)$ denote the connected component associated with
name $i$. The abundance and accumulated prestige of name $i$ are then
defined as
\begin{equation}
n_i(t)=|C_i(t)|,
\qquad
x_i(t)=\sum_{v\in C_i(t)} \eta_v ,
\end{equation}
and the total prestige in the system is
\begin{equation}
X(t)=\sum_i x_i(t)=\sum_{v} \eta_v .
\end{equation}

At each time step, a new node $v_t$ enters the system with intrinsic prestige
$\eta_{v_t}=K$, where $K\sim P(K)$. With probability $\rho(t)$, the
new node does not attach to any previous node. It therefore becomes the root of
a new connected component, corresponding to the introduction of a new name:
\begin{equation}
n_{\mathrm{new}}(t+1)=1,
\qquad
x_{\mathrm{new}}(t+1)=K .
\end{equation}

With complementary probability $1-\rho(t)$, the new node attaches to one
previous node $u$. The target node is selected with probability proportional
to its intrinsic prestige,
\begin{equation}
\Pi_u(t)=\frac{\eta_u}{\sum_{w} \eta_w}
=\frac{\eta_u}{X(t)} .
\end{equation}

If $u\in C_i(t)$, the new node joins component $C_i$. Therefore,
\begin{equation}
n_i(t+1)=n_i(t)+1,
\qquad
x_i(t+1)=x_i(t)+K .
\end{equation}

\subsection{Coarse-graining: from nodes to names}

The probability that a reused name is $i$ is the probability that the new node
attaches to any node belonging to component $C_i(t)$. Hence,
\begin{equation}
P_i(t)
=
[1-\rho(t)]\sum_{u\in C_i(t)} \Pi_u(t)
=
[1-\rho(t)]\sum_{u\in C_i(t)}
\frac{\eta_u}{X(t)} .
\end{equation}
Using the definition of component prestige,
\begin{equation}
\sum_{u\in C_i(t)}\eta_u=x_i(t),
\end{equation}
we obtain
\begin{equation}
P_i(t)
=
[1-\rho(t)]\frac{x_i(t)}{X(t)} ,
\end{equation}
which is exactly the reuse probability of the RIP model. Similarly, the
probability of creating a new component is
\begin{equation}
P_{\mathrm{new}}(t)=\rho(t).
\end{equation}

Thus, after coarse-graining connected components into names, the growing network
is pathwise equivalent to the RIP urn process.

\subsection{Relation to standard network growth models}

This representation differs from classical preferential-attachment models.
In the Barabási--Albert model \cite{barabasi1999emergence}, attachment is driven by degree, whereas here the
target node is selected according to its intrinsic prestige and independently
of its degree. A node with small degree can therefore attract later rulers if
its prestige is sufficiently large. The model also differs from the
Bianconi--Barabási model \cite{bianconi2001bose}, where attachment depends on a product of degree and
fitness. Here, the microscopic attachment rule is purely fitness-driven.

coarse-graining: names with larger accumulated prestige contain more total
fitness and are therefore more likely to attract future nodes. Innovation
corresponds to the birth of new connected components. The RIP model is therefore
equivalent to a growing fitness forest with component birth, where component
sizes are name abundances and component weights are accumulated name prestige.

\subsection{Realizations of the isomorphic growing network model}

In Supplementary Fig.~\ref{fig:popes-art} we show network visualizations corresponding to one realization of the RIP model for each ruling dynasty. The parameters of the prestige-increment distribution are those reported in Table~1. Nodes represent individual rulers or reigns, connected components represent names, and node size is proportional to individual prestige. In Supplementary Fig.~\ref{fig:distributions} we further quantify these network structures through their degree distributions $P(k)$. The datasets for which prestige is required display broader degree distributions, indicating the emergence of hubs generated by rare high-prestige reigns. In contrast, datasets well explained by random sampling dynamics show steeper degree profiles, consistent with all nodes having the same prestige.

\begin{figure*}[h]
  \centering
  \includegraphics[width=0.95\linewidth]{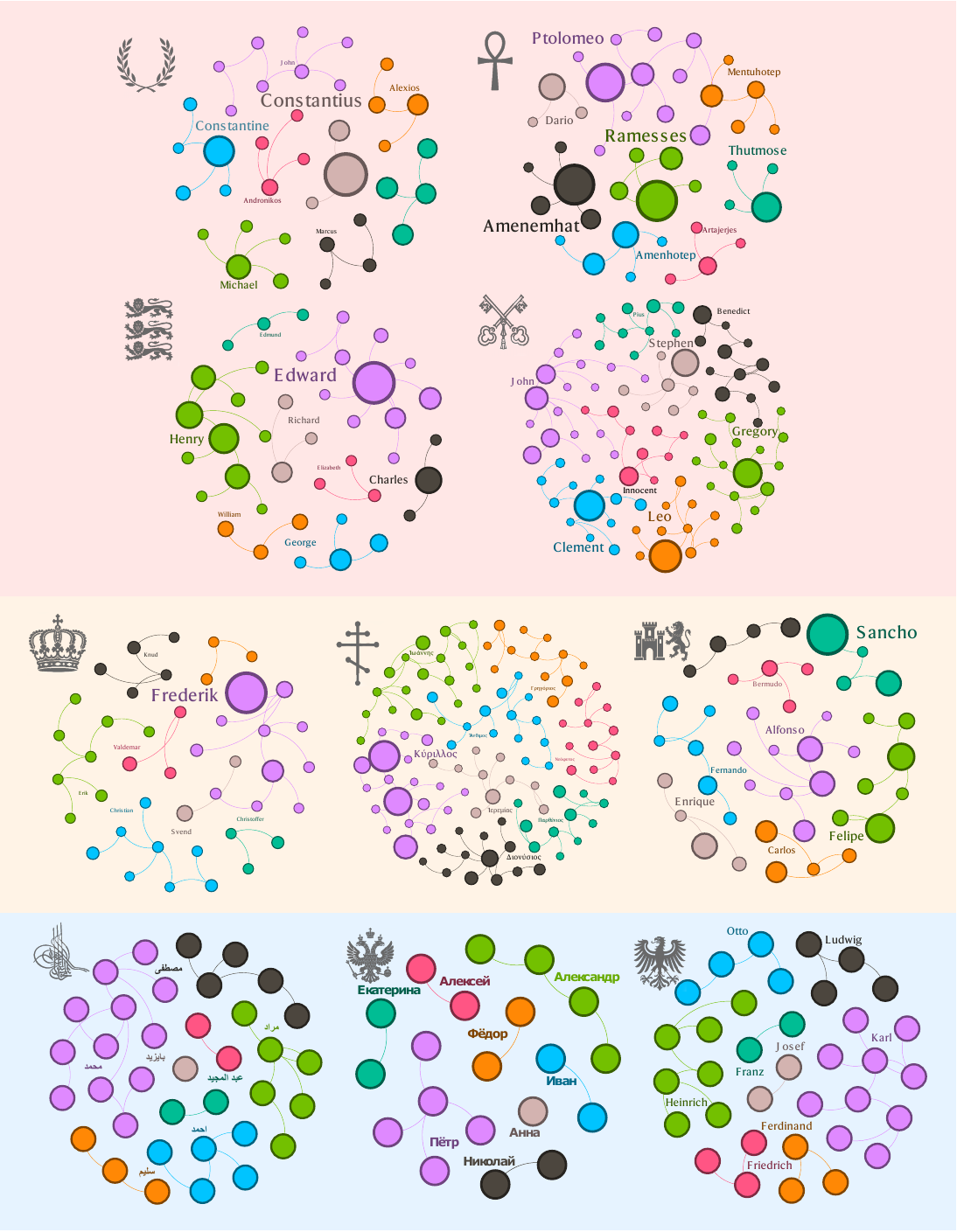}
\caption{\justifying \textbf{Visual representation of the growing network model isomorphic to the RIP model.} Nodes represent individual rulers. Edges connect a new ruler to the specific predecessor selected with probability proportional to its individual prestige, rather than to the most recent usage or the name's founder. Consequently, distinct connected components emerge naturally as lineages of names, where the size of each component corresponds to the name's total frequency.}
  \label{fig:popes-art}
\end{figure*}

\begin{figure*}[h]
  \centering
  \includegraphics[width=1\linewidth]{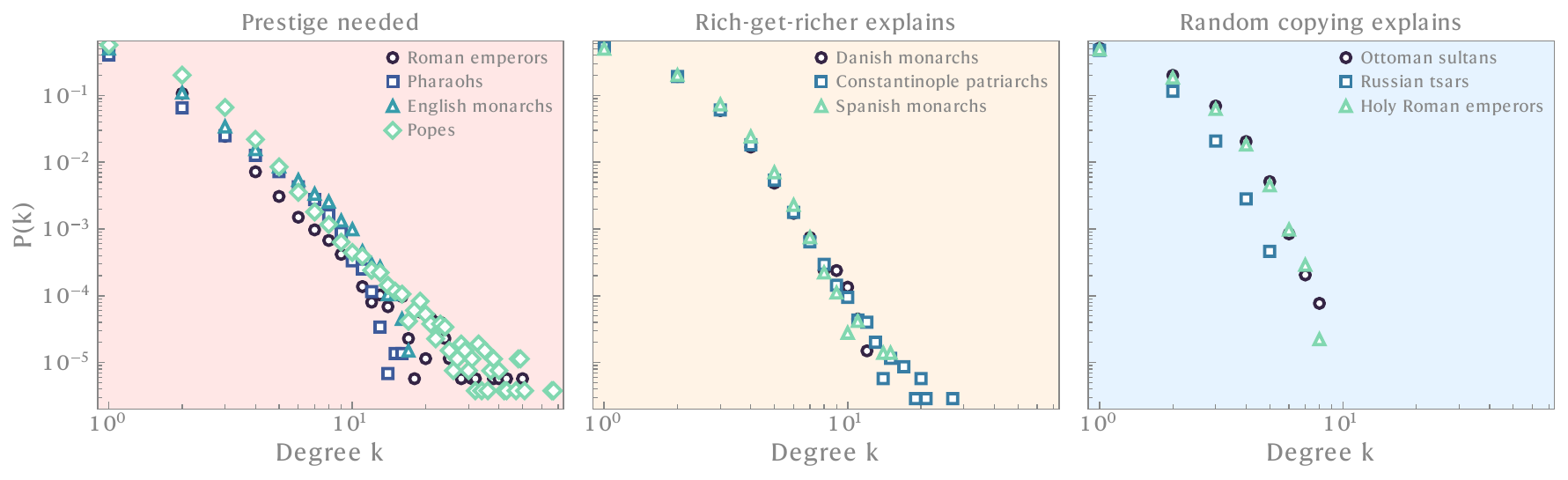}
\caption{\justifying \textbf{Degree distributions of the network representation of the RIP model.}
Degree distributions $P(k)$ of the RIP growing network representation for the ten dynastic naming traditions, grouped according to the mechanism identified in Fig.~2 of the main text. For each dataset, the distribution is averaged over $10^3$ independent realizations using the empirical innovation function $\rho(t)$ and the fitted prestige-increment distribution $P(K)$.}
  \label{fig:distributions}
\end{figure*}

\clearpage

\end{document}